\documentclass[aps,prd,twocolumn,showpacs,nofootinbib]{revtex4-2}

\usepackage{amssymb}
\usepackage{amsmath}
\usepackage{graphicx}
\usepackage{dcolumn}
\usepackage{bm}
\usepackage{epsfig}
\usepackage{amssymb}
\usepackage{color}
\usepackage{datetime}
\usepackage{wasysym}
\usepackage{slashed}
\usepackage[mathscr]{euscript}
\usepackage{hyperref}
\usepackage{tikz}
\usetikzlibrary{snakes}

\newcommand{\bal}{\begin{align}}
\newcommand{\eal}{\end{align}}
\newcommand{\beq}{\begin{eqnarray}}
\newcommand{\eeq}{\end{eqnarray}}
\newcommand{\nneeq}{\nonumber \end{eqnarray}}

\newcommand{\es}{& = &}

\newcommand{\ul}{  }

\newcommand{\cM}{ {\cal M} }

\newcommand{\cL}{ {\cal L} }

\renewcommand{\vec}{\mathbf}

\begin{document}
\title{Weak transition form factors of heavy-light pseudoscalar mesons\\ for space- and timelike momentum transfers}

\author{Oliver Heger}
\email{oliver.heger@ilf.com}
\affiliation{ILF Consulting Engineers Austria GmbH, A-8074 Graz, Austria}

\author{Mar\'ia G\'omez-Rocha }
\email{mgomezrocha@ugr.es}
\affiliation{Departamento  de  F\'isica  At\'omica, Molecular y Nuclear\\ and Instituto Carlos I de F\'isica Te\'orica y Computacional\\  Universidad  de  Granada,  E-18071  Granada,  Spain}

\author{Wolfgang Schweiger }
\email{wolfgang.schweiger@uni-graz.at}
\affiliation{Institute of Physics, University of Graz, A-8010 Graz, Austria}

\date{\today }

\begin{abstract}
This paper deals with the description of weak $B^-\rightarrow D^0, \pi^0$ and $D^-\rightarrow {K}^0, \pi^0$ transition form factors in both, the space- and time-like momentum transfer regions, within a constituent-quark model. To this aim neutrino-meson scattering and semileptonic weak decays are formulated within the framework of point-form relativistic quantum mechanics to end up with relativistic invariant process amplitudes from which meson transition currents and form factors are extracted in an unambiguous way. For space-like momentum transfers, these form factors depend on the frame in which the $W M M^\prime$ vertex is considered. On physical grounds such a frame dependence is expected from a pure valence-quark picture, since a complete, frame independent description of form factors is supposed to require valence as well as non-valence contributions. Non-valence contributions, the most important being the $Z$-graphs, are, however, suppressed in the infinite-momentum frame ($q^2<0$). On the other hand, they can play a significant role in the Breit frame ($q^2<0$)  and in  the direct decay calculation ($q^2>0$), as a comparison with the infinite-momentum-frame form factors (analytically continued to $q^2>0$) reveals. Numerical results for the analytically continued infinite-momentum-frame form factors are found to agree very well with lattice data in the time-like momentum transfer region and also the experimental value for the slope of the $F^+_{B\rightarrow D}$ transition form factor at zero recoil is reproduced satisfactorily. Furthermore, these predictions satisfy heavy-quark-symmetry constraints and their $q^2$ dependence is well approximated by a pole fit, reminiscent of a vector-meson-dominance-like decay mechanism. We discuss how such a decay mechanism can be accommodated within an extension of our constituent-quark model, by allowing for a non-valence component in the meson wave functions, and we also address the question of wrong cluster properties inherent in the formulation of relativistic quantum mechanics employed in this article.
\end{abstract}

%\keywords{}
\maketitle

\section{Introduction}
This paper continues previous work in which semileptonic weak $B\rightarrow D, \pi$ and $D\rightarrow K, \pi$ decays \cite{GomezRocha:2012zd,Gomez-Rocha:2014aoa,Gomez-Rocha:2012wqk,Gomez-Rocha:2011jnc} have been investigated within a constituent-quark model making use of the point-form of relativistic quantum mechanics~\cite{Dirac:1949cp,Keister:1991sb,Biernat:2010tp,Kli:2018}. In this paper we study $B\rightarrow D, \pi$ and $D\rightarrow K, \pi$ transition form factors for space- and time-like momentum transfers, as can be measured in neutrino scattering and semileptonic weak decays. Relativistic invariance of our approach is guaranteed by means of the, so called, \lq\lq Bakamjian-Thomas construction\rq\rq ~\cite{Bakamjian:1953kh}. Starting from a multichannel mass operator with an instantaneous confining interaction between the quarks and appropriate vertex interactions for the coupling of the different channels, one ends up with a relativistic invariant amplitude. The weak four-vector hadron current can then be extracted from this amplitude in a unique way and the hadron transition form factors are obtained by the covariant decomposition of this current.

The same kind of approach has also been applied to calculate electromagnetic form factors of $\pi$~\cite{Biernat:2009my}, $\rho$~\cite{Biernat:2014dea}, $B$ and $D$ mesons~\cite{GomezRocha:2012zd} for space-like momentum transfers. There it turned out that, although the amplitude for electron meson scattering was relativistic invariant, the extracted electromagnetic hadron current did not have all the properties one would expect. It is a four-vector current, but supposedly due to wrong cluster properties, inherent in the Bakamjian-Thomas construction, the hadron current is not just a function of  in- and outgoing hadron momenta, but exhibits also a slight dependence on the momenta of the in- and outgoing electron. This spurious dependence on the electron momenta shows up in the covariant decomposition of the current, which requires also covariants that involve the electron four-momenta. Furthermore, a dependence of the form factors on Mandelstam $s$, the invariant mass squared of the whole electron-meson system, is observed. This $s$-dependence can be reinterpreted as a dependence on the frame in which the $\gamma^\ast M\rightarrow M$ subprocess is considered. Such a frame dependence of the form factors is not just a speciality of our approach, but is rather a general property of any attempt to calculate hadron form factors within constituent-quark models using just a one-body current. As we will show in the following discussion, one would even expect it on physical grounds if one tries to extract the form factors from a pure one-body current involving only valence degrees-of-freedom.

\begin{figure*}[t!]
\includegraphics[width=0.7\textwidth]{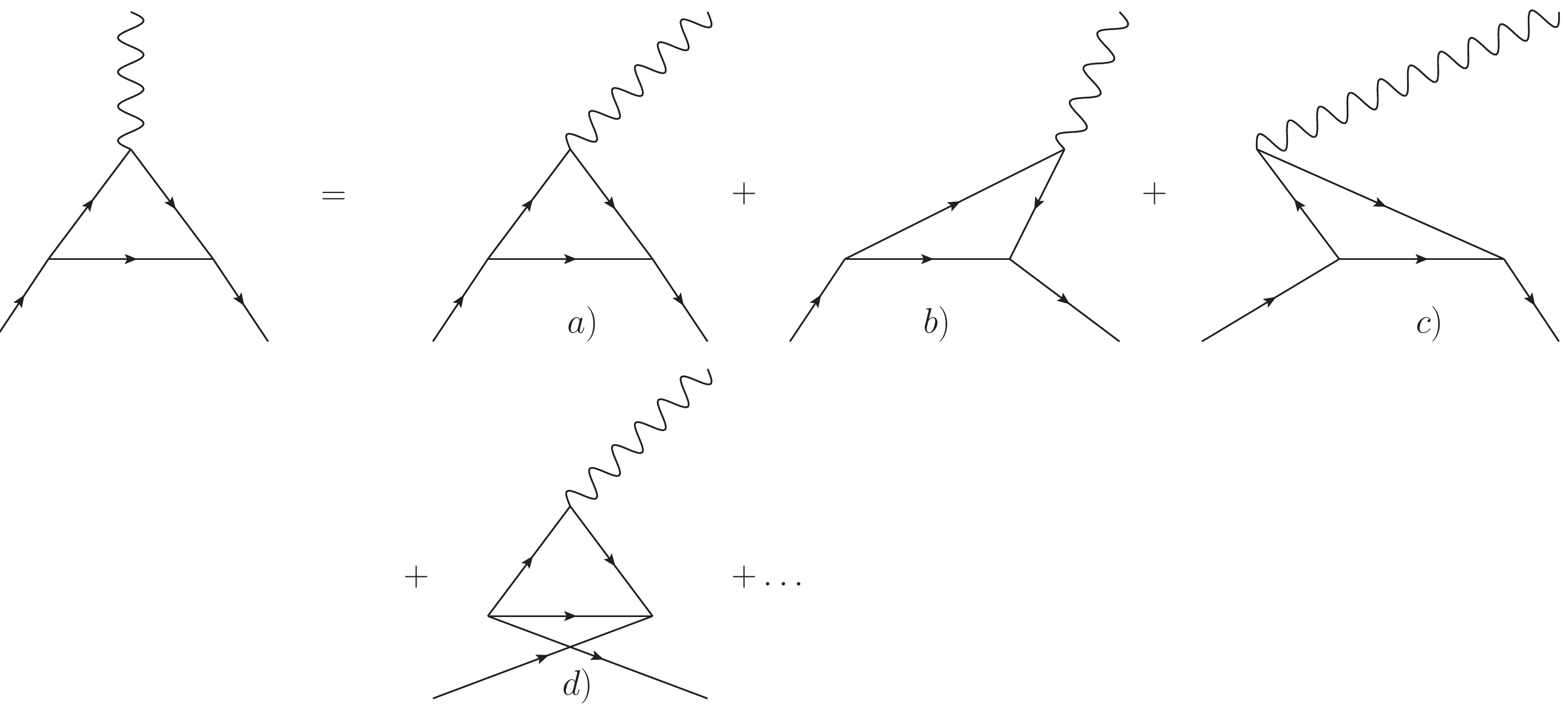}
\caption{ Decomposition of the covariant triangle diagram into time-ordered contributions. Time is running from  left to right.}\label{fig:triangle}
\end{figure*}

Several papers address the dependence of elastic and transition form factors of hadrons on the used frame and on the current components from which they are extracted (see, e.g., Refs.~\cite{Cheng:1997,Carbonell:1998rj,Simula:2002vm,Bakker:2003up,Li:2019kpr}).  These analyses make use of front-form dynamics and the main conclusion is that the construction of a frame independent covariant meson (transition) current requires the inclusion of the, so called, \lq\lq $Z$-graph\rq\rq\ contribution. This can already be seen in a simple $\phi^3$ field-theoretical model by calculating the electromagnetic current and form factor from the Feynman triangle diagram and decomposing this (covariant) diagram into time ordered contributions, as is graphically done in Fig.~\ref{fig:triangle}. The relative importance of the different time-ordered contributions will then depend on the reference frame in which they are calculated and also on the form of relativistic dynamics which is employed.\footnote{An instructive example for the interplay of the two time ordered contributions occurring in leading-order two-particle scattering within a simple scalar model  is given in Ref.~\cite{Ji:2012ux}. There, variations of the reference frame and the form of dynamics are considered.} In front form, e.g., only the contributions $a)$ and $b)$ survive, since a massive particle-antiparticle pair cannot be created out of the vacuum due to three-momentum conservation at the vertex. It is, in particular, conservation of the $P^+$ component, that is always positive for massive particles, which forbids the creation of massive particles out of the vacuum. In the $q^+=0$ Drell-Yan-West frame, in which no momentum is transferred in longitudinal direction, even graph $b)$ -- this is the one usually termed \lq\lq $Z$-graph\rq\rq -- becomes negligible in the $J^+$ current component, which is usually used for the extraction of the form factors~\cite{Brodsky:1998hn}. The role of $Z$-graph contributions, however, increases the more momentum is transferred in longitudinal direction (see, e.g.,\cite{Simula:2002vm,Bakker:2003up,Li:2019kpr}). A special $q^+=0$ frame is the infinite-momentum frame (IMF), in which the incoming and outgoing hadron moves with large momentum into a fixed direction and momentum is transferred transverse to this direction. It is this particular frame which also allows to connect front-form with instant-form or point-form calculations. In Ref.~\cite{Biernat:2010tp}, e.g., it has been shown that the analytical expression for the electromagnetic pion form factor, derived within the point form in the IMF, is connected with the corresponding front-form expression by a simple change of momentum variables. Like in front form, only graph $a)$ survives, if IMF kinematics is used to calculate the electromagnetic current in instant or point form~\cite{Close:1979}. The reason is simply that in all other graphs at least one constituent has to move into a direction opposite to the one of the incoming particle. If the momentum of the incoming (or outgoing) particle goes to infinity, the probability for finding a constituent that moves in opposite direction to its parent particle vanishes.

\begin{figure*}[t!]
\includegraphics[width=0.7\textwidth]{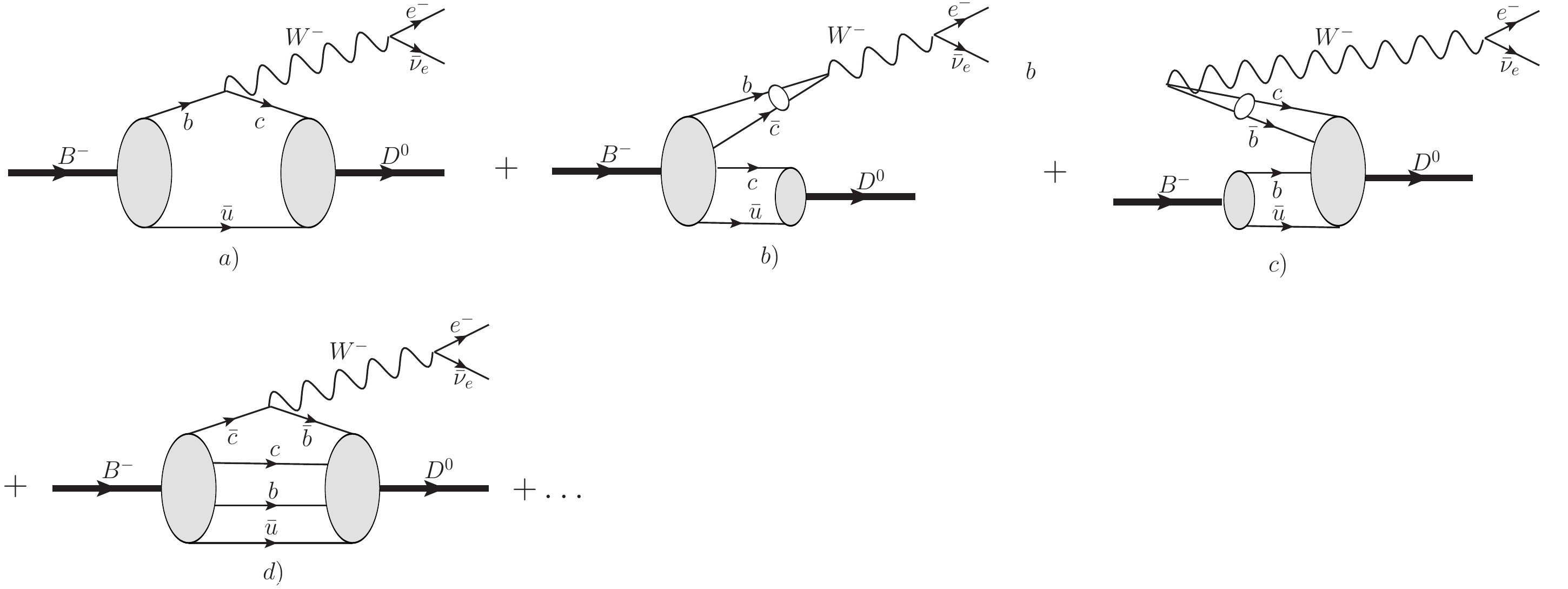}
\caption{ Possible mechanisms for the semileptonic $B^-\rightarrow D^0\, e^- \bar{\nu}_e$ decay within a constituent-quark model, allowing for a non-valence $\bar{q}Q\bar{Q}^\prime Q^\prime$ Fock component in the initial and final state. Diagrams $a)$, $b)$, $c)$ and $d)$ are the counterparts to the corresponding time-ordered diagrams in the $\phi^3$ model shown in Fig.~\ref{fig:triangle}.}\label{fig:CQMtriangle}
\end{figure*}

If one, however, is interested in form factors for time-like momentum transfers, as measured in decay processes, the decay kinematics requires $q^+>0$. This means that also the $Z$-graph $b)$ will provide a non-vanishing contribution to $J^+$ and the resulting form factor for time-like momentum transfers, if the calculation is done in front form~\cite{Cheng:1997,Simula:2002vm,Bakker:2003up,Li:2019kpr}. In instant and point form all possible time orderings are, in principle, necessary for time-like momentum transfers to end up with the complete, covariant triangle diagram. In a constituent-quark model the direct diagram $a)$ corresponds to the usual valence-quark contribution, whereas the other diagrams, in particular the $Z$-graph $b)$, are associated with higher Fock states.  This is graphically represented in Fig.~\ref{fig:CQMtriangle} for the semileptonic $B^-\rightarrow D^0\, e^- \bar{\nu}_e$ decay. The $Z$-graph contribution to this decay connects the $b\bar{u}c\bar{c}$ non-valence Fock state of the $B^-$ with the $c\bar{u}$ valence Fock state of the $D^0$. Graph $c)$ connects the $b\bar{u}$ valence Fock state of the $B^-$ with the $c\bar{u}b\bar{b}$ non-valence Fock state of the $D^0$. All the remaining graphs involve non-valence Fock states of $B^-$ and of $D^0$. This means that the $Z$-graph $b)$ provides presumably the most important non-valence contribution to the current. All other non-valence contributions are most likely much smaller, since they involve Fock states with an additional  $b\bar{b}$ pair which have tiny probability.

But let us now have a closer look at the $Z$-graph $b)$. The quarks in the intermediate $b\bar{u}c\bar{c}$ state are subject to a confining interaction. In the simplest case the confining interaction acts between $c$ and $\bar{u}$ to give the $D^0$ and between $b$ and $\bar{c}$ to give the series of $B_c^\ast$ resonances which can fluctuate into $W^-$. This leads to a vector-meson-dominance-like picture for the decay process and associated poles in the form factors which are located at $q^2=m_{B_c^\ast}^2$, i.e. at unphysical values of the time-like momentum transfer. The most important pole is the one closest to the zero-recoil point $q^2_{\mathrm{max}}=(m_B-m_D)^2$, where $q^2$ reaches its  maximum value. It comes from the $B_c^\ast$ ground state. This pole is still far away from $q^2_{\mathrm{max}}$, so that one would not expect substantial effects on the form factors. For the $B\rightarrow \pi$  decay, however, the nearest pole comes from $B^\ast$. It is already much closer to the zero-recoil point and hence the $Z$-graph contribution to the form factors gains importance as compared to the valence contribution~\cite{Isgur:1989qw}.

In the present paper we will not follow the strategy to calculate the $Z$-graph contribution explicitly, but we will rather try to estimate it and see whether our estimate exhibits a monopole-like behavior, as one would expect from a vector-meson-dominance-like mechanism. The idea is to derive analytical expressions for the weak transition form factors for space-like momentum transfers in the IMF, where the $Z$-graph is suppressed, and continue these form factor expressions analytically to time-like momentum transfers. Provided the analytical continuation is done correctly, this should already give the complete transition form factors for the decay. We can then compare the analytic continuation of the IMF result with the outcome of a direct decay calculation in which just the valence contribution is taken into account. Finally we check, whether the differences between these two ways to compute the decay form factors exhibit approximately a monopole-like $q^2$ dependence which would be typical for the missing $Z$-graph contribution in the direct decay calculation.

The general relativistic framework, in particular the point-form version of the Bakamjian-Thomas construction and the velocity-state representation, is summarized in Sec.~\ref{sec:general}. Section~\ref{sec:scatt} is devoted to the multichannel formulation of neutrino-meson scattering, $\nu_e M\rightarrow e^- M^\prime$, within our constituent-quark model, the extraction of the meson transition current from the scattering amplitude and its covariant decomposition which leads to weak transition form factors for space-like momentum transfers. Using the same model, the amplitude for the semileptonic $M\rightarrow M^\prime e^- \bar\nu_e$ decay is derived in Sec.~\ref{sec:decay} and the form factors for time-like momentum transfers are extracted from the weak transition current. Numerical studies of weak transition form factors for $B^-\rightarrow D^0$, $B^-\rightarrow \pi^0$, $D^-\rightarrow K^0$ and $D^-\rightarrow \pi^0$ are presented in Sec.~\ref{sec:numerics}. This starts with the introduction of the kinematics. In the following the frame dependence of the form factors is studied for space-like momentum transfers by comparing IMF results with Breit frame (BF) results. The difference gives already a clue of the possible size of the $Z$-graph contribution for space-like momentum transfers in frames different from the IMF. The space-like form factors are then analytically continued to time-like momentum transfers and compared with the form factors obtained from the decay amplitude. This is first done for BF kinematics, which is closest to the decay kinematics, to check the analytic continuation procedure. The comparison of the analytically continued IMF form factors with the decay form factors will then give us an estimate of the $Z$-graph contribution to the time-like form factors in the rest frame of the decaying particle. In the sequel we will check whether this difference exhibits a monopole-like behavior so that it can be ascribed to a missing  non-valence $Z$-graph contribution in the decay calculation. In Sec.~\ref{sec:HQL} we consider the heavy-quark limit and convince ourselves that the form factors tend to the Isgur-Wise function and that the differences between the analytically continued IMF form factors and the decay form factors vanish. This indicates that the $Z$-graph is suppressed in the heavy-quark limit, as one would expect. Section~\ref{sec:conclusions} contains our conclusions and an outlook.

\section{General framework}\label{sec:general}

For a proper relativistic description of the weak transition processes we are interested in, we make use of the Bakamjian-Thomas construction~\cite{Bakamjian:1953kh} and choose the point form of relativistic dynamics~\cite{Dirac:1949cp}. The point form has the feature that, from the 10 generators of the Poncar\'e group, those for the entire Lorentz subgroup are kinematic (i.e. these generators do not contain interactions), whereas the four components of the four-momentum operator $\hat P^\mu$ are dynamic (i.e. contain interaction terms). One main advantage of using the point form is that boosts of wave functions are simpler than in other forms of dynamics and that the addition of angular momenta is also facilitated.

In the point-form version of the Bakamjian-Thomas construction the 4-momentum operator is factorized into an interaction-dependent mass operator and a free 4-velocity operator~\cite{Keister:1991sb}
\begin{equation}\label{eq:massop}
\hat{P}^{\mu}=\hat{\mathcal M}\,  \hat V^{\mu}_{\mathrm{free}}=
\left(\hat{\mathcal M}_{\mathrm{free}}+ \hat{\mathcal
M}_{\mathrm{int}} \right) \hat V^{\mu}_{\mathrm{free}}\, .
\end{equation}
In this way the dynamics of the system is completely encoded in the mass operator.

Creation and annihilation of particles is described by means of a coupled-channel framework. This means that the mass operator $\hat \cM$ acts on a direct sum of multiparticle Hilbert spaces. The diagonal matrix elements of $\hat \cM$ are the sums of the relativistic kinetic energies of the particles in the corresponding channel plus instantaneous interactions between them (if present), like the quark-antiquark confinement potential. The off-diagonal matrix elements of $\hat \cM$ are vertex operators $\hat K$,  $\hat K^\dagger$ which describe the absorption or emission of particles, giving rise  to the transition from one channel to another. These vertex operators can be constructed starting from common field theoretical interaction-Lagrangean densities~\cite{Klink:2000pp}.

In this framework the most convenient representation of the Poincar\'e algebra is accomplished by means of \textit{velocity states}~\cite{Klink:1998zz}. These are multiparticle momentum states in the rest frame,
\begin{eqnarray}
|\vec k_i , \mu_i \rangle
\equiv
|\vec k_1 , \mu_1 ; \vec k_2 , \mu_2 ; ... ; \vec k_n , \mu_n \rangle  \ ,
\end{eqnarray}
with $\sum_{i=1}^n \vec{k}_i = 0 $ and $\mu_i$ being the $z$-projection of the (canonical) spin, which are boosted to an overall 4-velocity $V^\mu$ (with $V^\mu V_\mu=1$) by means of a rotationless (canonical) boost $B_c(V)$
\begin{eqnarray}
|V ;\vec k_1 , \mu_1 ; \vec k_2 , \mu_2 ; ... ; \vec k_n , \mu_n \rangle \\
&& \hspace{-1.5cm}:=
\hat U_{B_c(V)}|\vec k_1 , \mu_1 ; \vec k_2 , \mu_2 ; ... ; \vec k_n , \mu_n \rangle  \ . \nonumber
\end{eqnarray}
Our vertex operators are obtained from usual field  theoretical Lagrangean densities.
This is done by relating the matrix elements of a vertex operator between velocity states to the respective interaction-Lagrangean density in the following way~\cite{Klink:2000pp}
\begin{eqnarray}
\langle V'; \vec k'_i , \mu'_i | \hat K |V; \vec k_i , \mu_i \rangle
&&\\ && \hspace{-1.5cm}= N V^0 \delta^3(\vec V - \vec V')
\langle \vec k'_i , \mu'_i | \hat \cL_{\text{int}} | \vec k_i , \mu_i \rangle \, . \nonumber
\end{eqnarray}
The normalization factor $N$ is determined by the normalization of the velocity states.
The velocity-conserving delta function guarantees  overall 4-velocity conservation at the interaction vertices and thus the Bakamjian-Thomas-type structure of the 4-momentum operator as given in Eq.~(\ref{eq:massop}). Explicit expressions for such vertex matrix elements for the coupling of a photon or $W$-boson can be be found in Refs.~\cite{Biernat:2009my} and \cite{GomezRocha:2012zd}, respectively.

\medskip

\section{ Neutrino-meson scattering }\label{sec:scatt}

\subsection{The 1$W$-exchange amplitude and the weak current}
\begin{figure}[t!]
\includegraphics[width=0.3\textwidth]{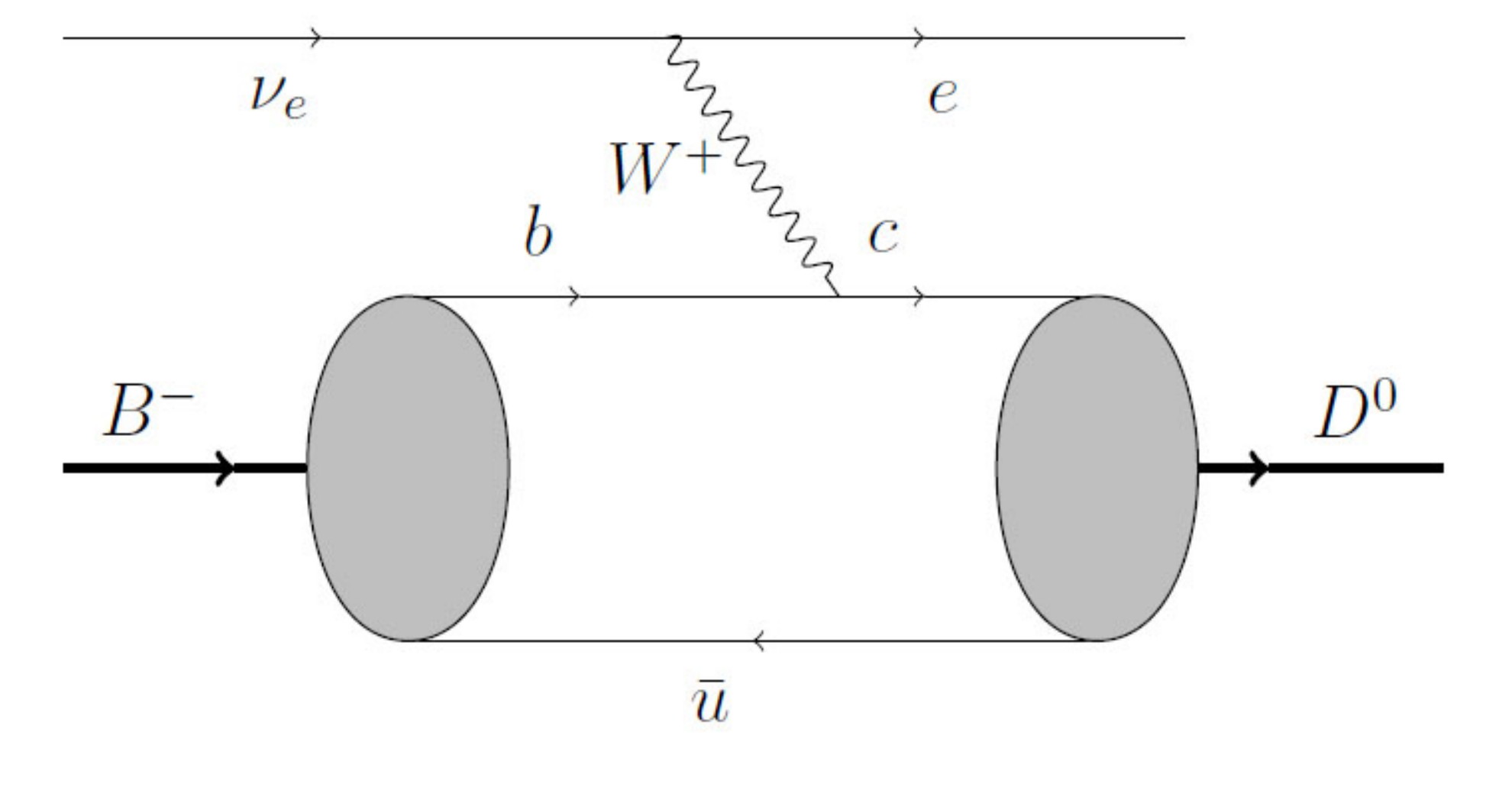}
\includegraphics[width=0.3\textwidth]{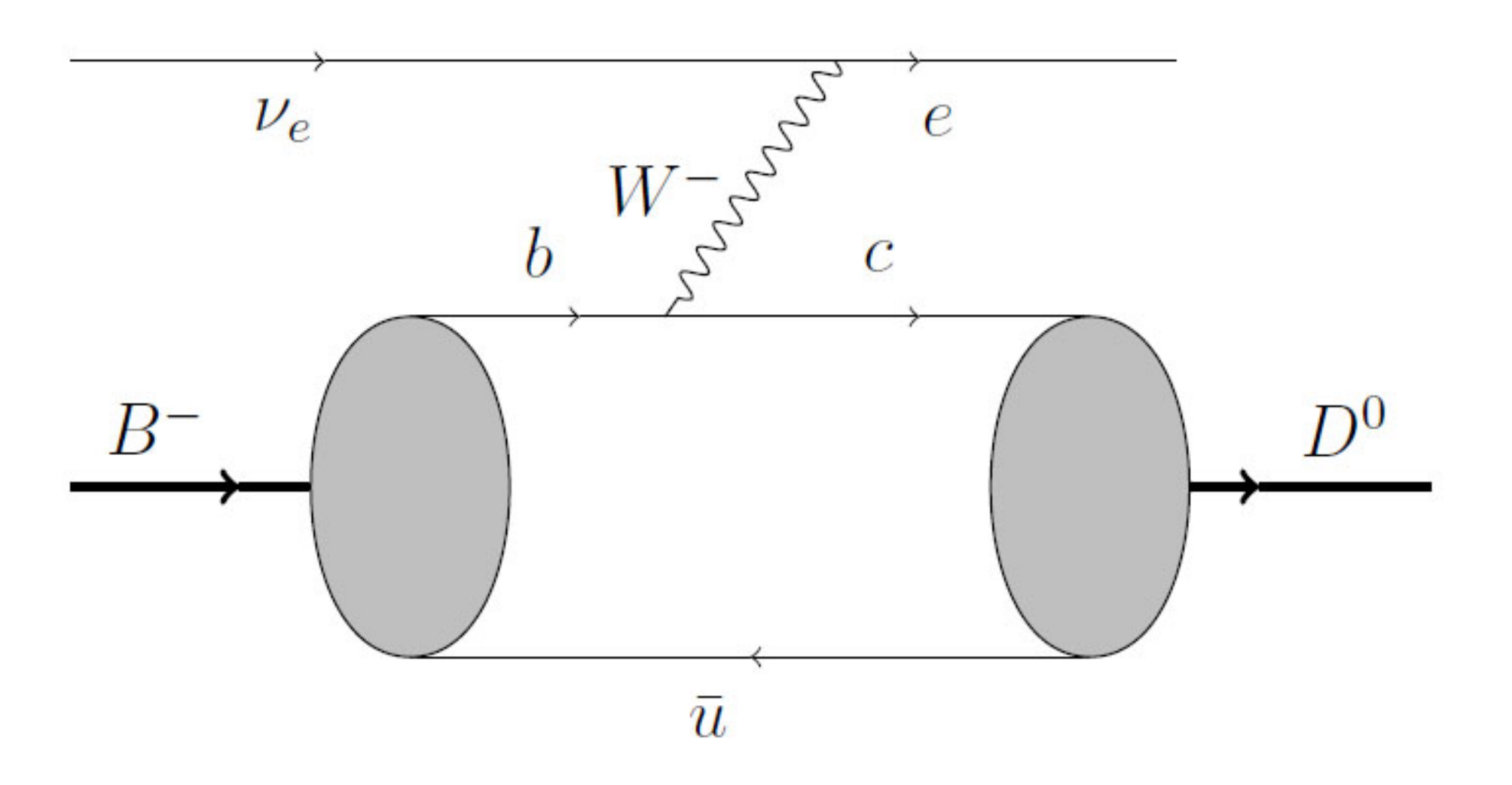}
\caption{ The two time orderings contributing to the invariant 1$W$-exchange amplitude for $\nu_e\, B^- \to e^-\, D^0$ scattering. }\label{fig:scattering}
\end{figure}

To extract the weak $M\rightarrow M^{\prime}$ transition current and the corresponding weak transition form factors for spacelike momentum tranfers, we first calculate the invariant 1$W$-exchange amplitude. Without loss of generality let us consider $\nu_e\,  B^- \rightarrow e^-\, D^0$ scattering,
where the $B^-$ consists of a heavy $b$ and a light $\bar{u}$ quark, whereas the $D^0$ contains a heavy $c$ and a light $\bar{u}$ quark in our constituent-quark picture.
%\footnote{ We emphasize that the presented derivations do not depend
%on the heavy-light structure of the outgoing meson and all calculations also hold for heavy-light to light-light transitions.}
The transition amplitude can be decomposed into the two time-ordered graphs shown in Fig.~\ref{fig:scattering}.

\begin{widetext}
The matrix mass operator, which is our starting point for the calculation of the 1-\textit{W}-exchange contribution to  $\nu_e\, B^- \to e^-\, D^0$ scattering, reads:
\begin{equation}\label{eq:mscatt}
 \hat{M}=\left( \begin{array}{cccc}
         \hat{M}^{\mathrm{conf}}_{\bar{u}b\nu_e} & \hat{K}_{\bar{u}b\nu_e\rightarrow \bar{u} b W e} & 0 & \hat{K}_{\bar{u}b\nu_e \rightarrow \bar{u}c W \nu_e} \\
         \hat{K}_{\bar{u}b\nu_e\rightarrow \bar{u} b W e}^{\dagger} & \hat{M}^{\mathrm{conf}}_{\bar{u}b W e} & \hat{K}_{\bar{u} c e \rightarrow  \bar{u} b W e}^\dagger& 0 \\
         0 & \hat{K}_{\bar{u} c e \rightarrow  \bar{u} b W e}& \hat{M}^{\mathrm{conf}}_{\bar{u} c e} &\hat{K}_{\bar{u}ce \rightarrow \bar{u} c W \nu_e} \\
         \hat{K}_{\bar{u}b\nu_e \rightarrow \bar{u}c W \nu_e}^{\dagger} & 0 & \hat{K}_{\bar{u}ce \rightarrow \bar{u} c W \nu_e}^{\dagger} & \hat{M}^{\mathrm{conf}}_{\bar{u}c W \nu_e} \\
        \end{array}
 \right)\, .
\end{equation}
\end{widetext}
The diagonal matrix elements contain the kinetic energies of the respective particles and an instantaneous confinement potential between the quark-antiquark pair, which is indicated by the label \lq\lq conf\rq\rq .
The resulting mass-eigenvalue equation is then
\begin{equation}
 \hat{M} \left( \begin{array}{l}
		    \lvert \Psi_{\bar{u}b\nu_e} \rangle \\
		    \lvert \Psi_{\bar{u}b W e} \rangle \\
		    \lvert \Psi_{\bar{u}c e} \rangle \\
		    \lvert \Psi_{\bar{u}c W \nu_e} \rangle \\
		 \end{array}
	\right)
= m  \left( \begin{array}{l}
		    \lvert \Psi_{\bar{u}b\nu_e} \rangle \\
		    \lvert \Psi_{\bar{u}b W e} \rangle \\
		    \lvert \Psi_{\bar{u}c e} \rangle \\
		    \lvert \Psi_{\bar{u}c W \nu_e} \rangle \\
		 \end{array}
	\right) \, ,
\end{equation}
where $m$ is the invariant mass of the whole system.
By means of a Feshbach reduction the channels containing the \textit{W} can be eliminated and one ends up
with an optical potential that describes the transition between the $\bar{u}b\nu_e$ and the $\bar{u} c e$ channel (for a graphical representation, see Fig.~\ref{fig:scattering}):
\begin{eqnarray}\label{eq:vopt}
\hat{V}_{\text{opt}}^{\bar{u}b \nu_e\rightarrow\bar{u}c e}(m)&=&\hat{K}_{\bar{u}ce \rightarrow  \bar{u}bWe}(m-\hat{M}^{\mathrm{conf}}_{\bar{u}bWe})^{-1}\hat{K}_{\bar{u}b\nu_e\rightarrow \bar{u}bWe}^{\dagger} \nonumber \\ && \hspace{-2.0cm}
\ + \ \hat{K}_{\bar{u}ce \rightarrow \bar{u}cW\nu_e}(m-\hat{M}^{\mathrm{conf}}_{\bar{u}cW\nu_e})^{-1}\hat{K}_{\bar{u}b\nu_e\rightarrow\bar{u}cW\nu_e}^{\dagger}.
\end{eqnarray}

Since we are only interested in the Born term for $\nu_e\,B^-\rightarrow e^-\, D^0$ scattering we just have to consider on-shell matrix elements ($m={k}_B^0+{k}_{\nu_e}^0={k}_D^0+{k}_e^0$) of
$\hat{V}_{\text{opt}}^{\bar{u}b \nu_e\rightarrow\bar{u}c e}$ between (velocity) eigenstates $\lvert {{V}};{\textbf{k}}_B;{\textbf{k}}_{\nu_e},{\mu}_{\nu_e} \rangle$ and $\lvert {{V}}^{\prime}; {\textbf{k}}_D; {\textbf{k}}_e,{\mu}_e \rangle$ of $\hat{M}^{\mathrm{conf}}_{\bar{u}b\nu_e}$ and $\hat{M}^{\mathrm{conf}}_{\bar{u}c e}$, respectively:
\begin{equation}
\label{scatteringamplitude}
\langle {{V}}^{\prime}; {\textbf{k}}_D; {\textbf{k}}_e,{\mu}_e \rvert
\hat{V}_{\text{opt}}^{\bar{u}b \nu_e\rightarrow\bar{u}c e}(m)
\lvert {{V}};{\textbf{k}}_B;{\textbf{k}}_{\nu_e},{\mu}_{\nu_e} \rangle_{\text{os}} \ .
\end{equation}
For better readability we have suppressed all discrete quantum numbers which are necessary to specify the meson states uniquely. Note that incoming and outgoing states contain $B$ and $D$ mesons, rather than quarks, since they are eigenstates of mass operators containing a confining interaction between quark and antiquark. Inserting pertinent completeness relations into Eqs.~(\ref{eq:vopt}) and (\ref{scatteringamplitude}) gives us an expression for the leading-order $\nu_e\,  B^- \rightarrow e^-\, D^0$ amplitude in terms of vertex matrix elements and the $\bar{q} Q$ wave functions (for a detailed derivation, see Ref.~\cite{GomezRocha:2012zd}). The resulting transition amplitude looks like the leading order perturbative result in quantum field theory. It is essentially the contraction of the leptonic current with a hadronic current, multiplied with the $W$-propagator:
\begin{widetext}
\begin{eqnarray}\label{eq:voptme}
\langle {{V}}^{\prime};{\textbf{k}}_D;{\textbf{k}}_e,{\mu}_e\rvert \hat{V}_{\text{opt}}^{\bar{u}b\nu_e\rightarrow \bar{u}ce}(m)
 \lvert {{V}};{\textbf{k}}_B;{\textbf{k}}_{\nu_e},{\mu}_{\nu_e}\rangle
&=&
 {V}^0 \delta^3(\mathbf{{V}}-\mathbf{{V}}^{\prime}) \frac{(2\pi)^3}{\sqrt{({k}_D^0+{k}_e^0)^3}\sqrt{({k}_B^0+{k}_{\nu_e}^0)^3}} \\
 && \hspace{-2.0cm}\times\frac{-e^2 V_{\text{cb}}}{2 \sin{\vartheta_W}^2} \frac{1}{2} \bar{u}_{\mu_e}(\mathbf{{k}}_e)\gamma^{\mu}(1-\gamma^5)u_{\mu_{\nu_e}}(\mathbf{{k}}_{\nu_e}) \, \frac{(-g^{\mu \nu}+\frac{k_W^{\mu}k_W^{\nu}}{m_W^2})}{q^2-m_W^2}
 \frac{1}{2} \tilde{J}_{B\rightarrow D}^{\nu}(\mathbf{{k}}_D;\mathbf{{k}}_B)\, . \nonumber
\end{eqnarray}
Here, $\vartheta_{\mathrm{w}}$ denotes the electroweak mixing angle, $e$ the elementary electric charge and $V_{cb}$ the CKM
matrix element occurring at the $Wbc$-vertex. The hadronic (transition) current, containing the $Q\bar{q}$ wave functions, results as
\begin{eqnarray}\label{eq:Jspsps}
 \tilde{J}_{B\rightarrow D}^{\nu}(\mathbf{{k}}_D;\mathbf{{k}}_B)
& = &  2 \sqrt{{k}_D^0{k}_B^0} \int{\frac{d^3\tilde{k}^\prime_{\bar{u}}}{2k_b^{0}} \,
  \sqrt{\frac{k_{\bar{u}}^0+k_b^{0}}{k_{\bar{u}}^{\prime 0}+k_c^{\prime 0}}}\, \sqrt{\frac{\tilde{k}_{\bar{u}}^{\prime 0}+\tilde{k}_c^{\prime 0}}{\tilde{k}_{\bar{u}}^{0}+\tilde{k}_b^{0}}}
\,\sqrt{\frac{\tilde{k}_{\bar{u}}^{0}\tilde{k}_b^{0}}{\tilde{k}_{\bar{u}}^{\prime 0}\tilde{k}_c^{\prime 0}}}} \nonumber \\
&&\times  \sum_{\mu_{\bar{u}} \mu_b \mu_c^\prime}{\bar{u}_{\mu_c^\prime}(\textbf{k}_c^\prime)\gamma^{\nu}(1-\gamma^5)u_{\mu_b}(\textbf{k}_b})
\Psi^{\ast}_{\mu_{\bar{u}}\mu_c^\prime}(\mathbf{\tilde{k}}_{\bar{u}}^\prime)
\Psi_{\mu_{\bar{u}}\mu_b}(\mathbf{\tilde{k}}_{\bar{u}})\, ,
\end{eqnarray}
\end{widetext}
where $\Psi_{\mu_{\bar{u}}\mu_b}(\mathbf{\tilde{k}}_{\bar{u}})$ and $\Psi^{\ast}_{\mu_{\bar{u}}\mu_c^\prime}(\mathbf{\tilde{k}}_{\bar{u}}^\prime)$ are the $\bar{q} Q$ wave functions of the incoming and outgoing $B$ and $D$ meson, respectively.
Since we  are dealing with velocity states, the momenta are CM momenta satisfying $\mathbf{k}_{\nu_e}+\mathbf{k}_B=\mathbf{k}_{\nu_e}+\mathbf{k}_{\bar{u}}+\mathbf{k}_b=\mathbf{k}_e+\mathbf{k}_D=\mathbf{k}_e+\mathbf{k}^{\,\prime}_{\bar{u}}+\mathbf{k}^{\,\prime}_c=0$. These relations, combined with the spectator condition $\vec{k}_{\bar{u}}=\vec{k}_{\bar{u}}^\prime$, imply in addition that the three-momentum transferred to the mesons and the heavy quarks is the same, $\vec{q}=\vec{k}_B-\vec{k}_D=\vec{k}_b-\vec{k}_c^\prime$. Furthermore, momenta with tilde satisfy $\tilde{\mathbf{k}}_{\bar{u}}+\tilde{\mathbf{k}}_{b}=\tilde{\mathbf{k}}_{\bar{u}}^\prime+\tilde{\mathbf{k}}_{{c}}^\prime=\vec{0}$. They are related to the momenta without tilde by means of canonical boosts, i.e. $k_i=B_c(v_{\bar{u} b}) \tilde{k}_i$, $i=\bar{u},b$ and
$k^\prime_i=B_c(v^\prime_{\bar{u} c}) \tilde{k}^\prime_i$, $i=\bar{u},c$, where $v^{(\prime)}_{\bar{q}Q}=(k_{\bar{q}}^{(\prime)}+k_Q^{(\prime)})/\sqrt{(k_{\bar{q}}^{(\prime)}+k_Q^{(\prime)})^2}$.

If the orbital part $\psi_{M}(\tilde{\mathbf{k}}_{\bar{q}})$ of the $B$ and $D$ meson wave function (at rest) is a pure $s$-wave, as we will assume for our numerical studies, the weak $B\rightarrow D$ transition current becomes
\begin{widetext}
\begin{eqnarray}
\label{pseudoscalarcurrent}
\tilde{J}_{B\rightarrow D}^{\nu}(\mathbf{{k}}_D;\mathbf{{k}}_B)\es  \frac{ \sqrt{{k}_D^0{k}_B^0}}{4 \pi} \int{\frac{d^3\tilde{k}^\prime_{\bar{u}}}{2k_b^{0}} \,
  \sqrt{\frac{k_{\bar{u}}^0+k_b^{0}}{k_{\bar{u}}^{\prime 0}+k_c^{\prime 0}}}\, \sqrt{\frac{\tilde{k}_{\bar{u}}^{\prime 0}+\tilde{k}_c^{\prime 0}}{\tilde{k}_{\bar{u}}^{0}+\tilde{k}_b^{0}}}
\,\sqrt{\frac{\tilde{k}_{\bar{u}}^{0}\tilde{k}_b^{0}}{\tilde{k}_{\bar{u}}^{\prime 0}\tilde{k}_c^{\prime 0}}}} \nonumber \\
&&\times  \sum_{\mu_b \mu_c^\prime}{\bar{u}_{\mu_c^\prime}(\textbf{k}_c^\prime)\gamma^{\nu}(1-\gamma^5)u_{\mu_b}(\textbf{k}_b})
\,\psi^{\ast}_{D}(|\mathbf{\tilde{k}}_{\bar{u}}^\prime|)
\,\psi_{B}(|\mathbf{\tilde{k}}_{\bar{u}}|) \nonumber \\
 && \times D_{\mu_b \mu_c^\prime}^{\frac{1}{2}}\left[{R}_W(\tilde{v}_b,B_c(v_{\bar{u} b})) \
  {R}_W^{-1}(\tilde{v}_{\bar{u}},B_c(v_{\bar{u}b}))
{R}_W(\tilde{v}^\prime_{\bar{u}},B_c(v^\prime_{\bar{u}c}))
 {R}_W^{-1}(\tilde{v}^\prime_c,B_c(v^\prime_{\bar{u}c}))\right].\nonumber\\
 \end{eqnarray}
 \end{widetext}
 The spin rotation matrix $\mathbf{D}^{1/2}$ with matrix elements $D_{\mu_b \mu_c^\prime}^{\frac{1}{2}}$ describes how  the spin orientation of the constituent quarks is affected by boosting the mesons from rest to their respective momenta $\mathbf{k}_B$ and $\mathbf{k}_D$. The argument of the spin rotation matrix is a series of Wigner rotations $R_W(v,\Lambda)=B_c^{-1}(\Lambda v) \Lambda B_c(v)$. For the practical calculation of $\mathbf{D}^{1/2}$ we refer to Ref.~\cite{Biernat:2011mp}. It should also be noted that $\tilde{J}_{B\rightarrow D}^{\nu}(\mathbf{k_D};\mathbf{k}_B)$ still does not transform like a four-vector under Lorentz transformations, but rather by means of a Wigner rotation~\cite{Biernat:2009my}. The reason is that we have been using velocity states for the derivation of the current (see Eq.~(\ref{eq:voptme})).  A four-vector current is obtained by going back to the physical meson momenta, $p_B=B_c(V)\, k_B$ and $p_D=B_c(V)\, k_D$, by means of a canonical boost $B_c(V)$ with the overall velocity $V$ of the electron-meson system. The resulting four-vector transition current then reads:
\begin{equation}\label{eq:Jcovps}
\tilde{J}^\nu_{B\rightarrow D}({p}_D;{p}_B)=B_c(V)^\nu_{~\mu}\, \tilde{J}_{B\rightarrow D}^{\mu}(\mathbf{{k}}_D;\mathbf{{k}}_B) \, .
\end{equation}
 It should be quite obvious, how all these considerations generalize to weak transitions involving mesons different from $B^-$ and $D^0$.

\enlargethispage{1\baselineskip}

\subsection{Weak transition form factors}

Having now an expression for the weak transition current between pseudoscalar quark-antiquark bound states in terms of constituent currents and bound-state wave functions, we can calculate the current numerically and extract weak transition form factors. Such form factors could, in principle, be measured in different neutrino-meson scattering processes like, e.g., $\nu_e \, B^-  \rightarrow e^-\,D^0$, $\nu_e\, B^- \rightarrow e^-\, \pi^0$, or $\nu_e\,D^- \rightarrow e^-\,\pi^0$. The general covariant structure of a  pseudoscalar weak transition current is determined by two covariants which can be built with $p_{M}$ and $p_{M^\prime}$,
the four-momenta of the incoming and outgoing mesons $M$ and $M^\prime$, respectively. Each covariant is multiplied with a Lorentz scalar which can only depend on the momentum transfer squared:
\begin{eqnarray}\label{decompositionone}
 J^{\nu}_{M\rightarrow M^\prime}(p_{M^{\prime}}; p_{M})&=&(p_{M}+p_{M^{\prime}})^{\nu} F^+(q^2)\nonumber\\&& + (p_{M}-p_{M^{\prime}})^{\nu} F^-(q^2).
\end{eqnarray}
An equivalent decomposition for such a current is given by \cite{Wirbel:1985ji}
\begin{eqnarray}
\label{decompositiontwo}
 J^{\nu}_{M\rightarrow M^\prime}(p_{M^{\prime}}; p_{M})&=& \frac{m_{M}^2-m_{M^{\prime}}^2}{q^2}q^{\nu}F_0(q^2)\\ &&\hspace{-1.0cm}+\left( (p_{M}+p_{M^{\prime}})^{\nu} - \frac{m_{M}^2-m_{M^{\prime}}^2}{q^2}q^{\nu}\right) F_1(q^2)\,.\nonumber
\end{eqnarray}
The four-momentum transfer is defined as $q=(p_{{M}}-p_{M^{\prime}})$ and $q^2=q^{\mu}q_{\mu}$.
By comparing both decompositions it follows immediately that
\begin{eqnarray}\label{eq:f0f1}
 F_1(q^2)&=&F^+(q^2).  \nonumber \\
 F_0(q^2)&=&F^+(q^2)+\frac{q^2}{m^2_M-m^2_{M^{\prime}}} F^-(q^2).
\end{eqnarray}
Here it is necessary to recall that the Bakamjian-Thomas construction spoils cluster separability.
As soon as a spectator is present in a Bakamjian-Thomas type mass operator, it affects interactions via the overall velocity-conserving delta function. This delta function is necessary for the factorization of the four-momentum operator
into the product of an interacting mass operator and a free four-velocity operator~\cite{Biernat:2011mp,Biernat:2010tp,Biernat:2014dea,GomezRocha:2012zd}. As a consequence, the current derived by means of our Bakamjian-Thomas type approach depends in one, or  another way also on the lepton momenta. For elastic scattering of an electron by a pseudoscalar meson it turned out that a complete covariant decomposition of the (conserved) current requires one additional spurious covariant, namely the sum of the incoming and outgoing electron four-momenta~\cite{Biernat:2009my,GomezRocha:2012zd}.\footnote{The occurrence of an additional covariant parallels the spurious dependence of currents on the orientation of the light front in the covariant light-front approach discussed in Ref.~\cite{Carbonell:1998rj}.}  In addition, the form factors are observed to be not only functions of the four-momentum transfer squared, but they depend also mildly on Mandelstam $s$, the invariant mass of the electron-meson system. For the weak pseudoscalar transition current there is no current-conservation condition and thus its covariant decomposition  involves already two covariants, the sum and the difference of incoming and outgoing meson momenta (see Eq.~(\ref{decompositionone})). These two covariants are also sufficient for a complete covariant decomposition of our model current. The dependence on the lepton momenta, however, still enters the form factors via the Mandelstam $s$ dependence. The general covariant decomposition of $ \tilde{J}^{\nu}_{M\rightarrow M^\prime}$, as given in Eqs.~(\ref{pseudoscalarcurrent}) and (\ref{eq:Jcovps}), is thus:
\begin{widetext}
\begin{equation}\label{eq:Jcovdec}
 \tilde{J}^{\nu}_{M\rightarrow M^\prime}(p_{M^{\prime}}; p_{M})= \left( (p_{M}+p_{M^{\prime}})^{\nu} - \frac{m_{M}^2-m_{M^{\prime}}^2}{q^2}q^{\nu}\right) F_1(q^2,s)+\frac{m_{M}^2-m_{M^{\prime}}^2}{q^2}q^{\nu}F_0(q^2,s)\, .
\end{equation}
\end{widetext}
The \textit{s}-dependence of the form factors may also be interpreted as a dependence on the frame in which the $W M \rightarrow M^{\prime}$ subprocess is considered.  In our case this frame dependence is well under control, since we observe that for large enough Mandelstam $s$ it becomes negligible. This suggests to extract the form factors in the limit $s\rightarrow\infty$. The corresponding kinematics corresponds to the IMF, in which the incoming and outgoing hadron moves with large momentum into a fixed direction and momentum is transferred by the $W$-boson transverse to this direction.
As already mentioned in the introduction, the IMF has also the advantage that the non-valence $Z$-graph contribution is suppressed. As long as one works within a valence-quark picture, it thus seems to be preferable to work with IMF kinematics when calculating hadron form factors. In reference frames different from the IMF, the $Z$-graph is likely to play a non-negligible role and a pure valence-quark description may miss part of the physics.

\section{ Semileptonic meson decays }\label{sec:decay}

Next we want to describe the semileptonic meson decay $M\rightarrow M^\prime\, e^- \bar{\nu}$ within our valence-quark picture, in order to extract the weak $M\rightarrow M^\prime$ transition form factors for timelike momentum transfers. Again, the Bakamjian-Thomas mass operator that yields the invariant amplitude
for the semileptonic weak decay of a meson $M$ into another meson $M^\prime$
requires four channels to allow for the two possible time orderings
of the $W$-boson exchange.  Without loss of generality we consider the $B^- \to D^{0} e \bar\nu_e$ transition. The
corresponding matrix mass operator has the form
\begin{widetext}
\begin{eqnarray}\label{eq:massopdecay}
\hat M \es \left(\begin{array}{cccc} \hat M^{\mathrm{conf}}_{\bar{u}b} & \hat{K}_{\bar{u}b\rightarrow \bar{u} c W}
&\hat{K}_{\bar{u}b\rightarrow \bar{u} b We\bar{\nu}_e} &0\\
\hat{K}^\dag_{\bar{u}b\rightarrow \bar{u} c W}& \hat M^{\mathrm{conf}}_{\bar{u} c W} &0&\hat{K}^\dag_{\bar{u} c e\bar{\nu}_e \rightarrow\bar{u}c W}  \\
\hat{K}^\dag_{\bar{u}b\rightarrow \bar{u} b We\bar{\nu}_e}&0&\hat M^{\mathrm{conf}}_{\bar{u} b We\bar{\nu}_e}&K^\dag_{\bar{u} c e\bar{\nu}_e  \rightarrow \bar{u} b We\bar{\nu}_e}\\
0&\hat{K}_{\bar{u} c e\bar{\nu}_e \rightarrow\bar{u}c W} & K_{\bar{u} c e\bar{\nu}_e  \rightarrow \bar{u} b We\bar{\nu}_e}& \hat M^{\mathrm{conf}}_{\bar{u} c e\bar{\nu}_e}
\end{array}\right)\, .
\end{eqnarray}
\end{widetext}
The label ``conf'' indicates again that an instantaneous confining potential between quark and anti-quark is included in the diagonal matrix elements.
\begin{figure}[b!]
\includegraphics[scale=0.5]{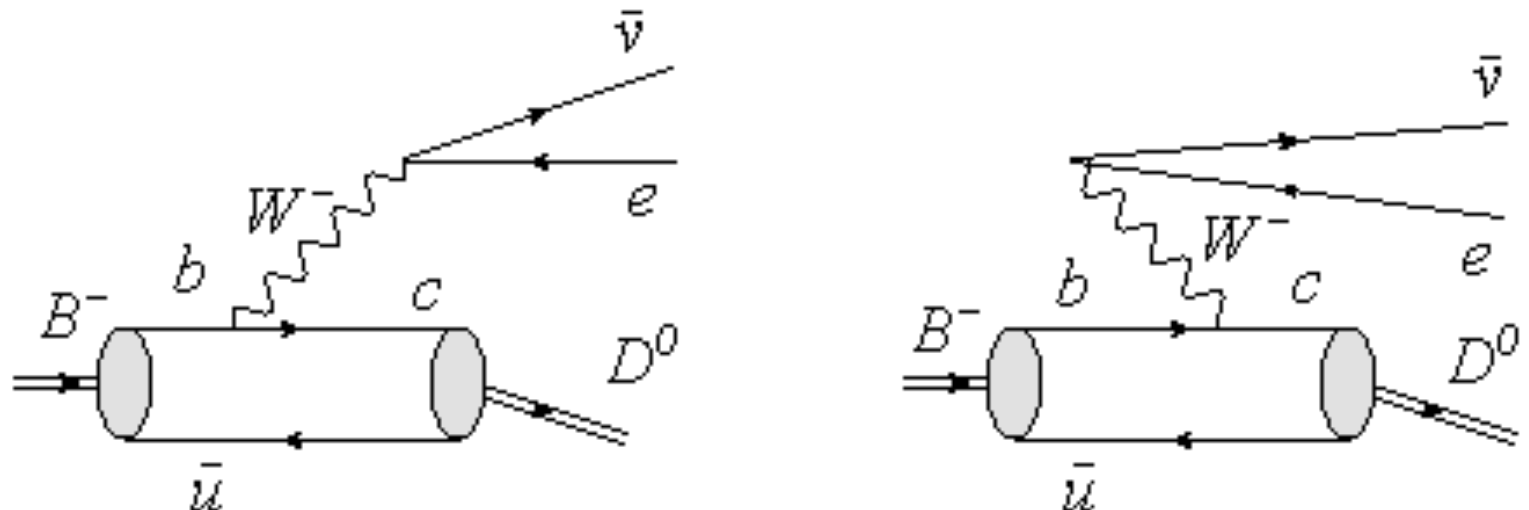}
\caption{ The two time orderings contributing to the invariant $1W$-exchange amplitude for $B^-  \to D^0 e \,\bar{\nu}_e$.}
\label{DecayGraph}
\end{figure}

The optical potential that describes the transition from the $\bar{u} b$ to the $\bar{u} c e \bar \nu_e$  channel is obtained by applying a
Feshbach reduction. This eliminates the $\bar{u} c W $ and $\bar{u} b W e \bar \nu_e$ channels. The transition potential has then two terms which correspond to the two possible time orderings (cf. Fig.~\ref{DecayGraph}). It reads:
\begin{eqnarray}\label{optweak}
\hat V_{\mathrm{opt}}^{\bar{u} b\rightarrow \bar{u} c e\bar{\nu}_e}(m)
&=&  \hat{K}_{\bar{u} c e\bar{\nu}_e \rightarrow \bar{u} c W}
(m-M^{\mathrm{conf}}_{\bar{u} c W})^{-1}\hat{K}^\dag_{\bar{u} b \rightarrow
\bar{u} c W}
\nonumber\\&&\hspace{-1.5cm} + \ \hat{K}_{\bar{u} c e\bar{\nu}_e \rightarrow
\bar{u} b W e \bar{\nu}_e} (m-\hat
M^{\mathrm{conf}}_{ \bar{u} b W e \bar{\nu}_e})^{-1}
 \hat{K}^\dag_{\bar{u} b\rightarrow \bar{u} b W e \bar{\nu}_e}\, .\nonumber\\
\end{eqnarray}

The weak hadronic current and, in the sequence, the decay form factors are extracted from
on-shell  matrix elements of $\hat V_{\mathrm{opt}}^{\bar{u} b\rightarrow \bar{u} c e\bar{\nu}_e}(m)$:
\begin{equation}\label{eq:vopttrans}
\langle {v}^\prime; \vec{{k}}_{D};
\vec{{k}}_e,
{\mu}_e; \vec{{k}}_{\bar{\nu}_e},\mu_{\bar{\nu}_e}
\vert \hat V_{\mathrm{opt}}^{\bar{u} b\rightarrow \bar{u} c e\bar{\nu}_e}(m)\vert \vec{{k}}_{B}
\rangle_{\mathrm{os}}\, .
\end{equation}
\lq\lq On-shell\rq\rq\ means $m=m_B = k_B^0 = k_D^0+k_e^0+k_{\bar{\nu}_e}^0$.  The analytical calculation of these on-shell matrix elements is explained in some detail in Refs.~\cite{Biernat:2009my,GomezRocha:2012zd,Gomez-Rocha:2013bga}, where also explicit expressions can be found. The final result for the invariant $B\to D$ decay amplitude looks again like
the one obtained from leading order covariant perturbation theory:
\begin{widetext}
\begin{eqnarray}\label{eq:voptcov}
&& \langle {v}^\prime;\vec{{k}}_{D}; \vec{{k}}_e,
{\mu}_e; \vec{{k}}_{\bar{\nu}_e}
\vert  \hat V_{\mathrm{opt}}^{\bar{u} b\rightarrow \bar{u} c e\bar{\nu}_e}(m) \vert \vec{{k}}_{B}=\vec{0}
\rangle_{\mathrm{os}}
\ = \
{v}_0\, \delta^3 (\vec{{v}}^{\, \prime} -
\vec{{v}}\, )\, \frac{(2 \pi)^3
}{\sqrt{(k^0_e+k^0_{\bar{\nu}_e}
+k^0_D)^3} \sqrt{
{k^0_B}^3}}
\nonumber
\\
&& \hspace{1.0cm} \times
\frac{e^2}{2\sin^2\vartheta_{\mathrm{w}}}V_{cb}\,
\frac{1}{2}\,{\bar{u}_{{\mu}_e}
(\vec{{k}}_e)\,\gamma^\mu (1-\gamma^5)
\, v_{{\mu}_{\bar{\nu}_e}}
(\vec{{k}}_{\bar{\nu}_e}) }
\ \frac{(-g_{\mu
\nu})}{({k}_e+{k}_{\bar{\nu}_e})^2-m_W^2}
\frac{1}{2}\,{\tilde{J}_{B\rightarrow
D}^\nu(\vec{{k}}_D;\vec{{k}}_B)}\, .
\end{eqnarray}
Assuming again that the orbital part of the $B$ and $D$ meson wave function (at rest) is a pure $s$-wave, the hadron current $\tilde{J}_{B\rightarrow D}^\nu(\vec{{k}}_D;\vec{{k}}_B)$ has the following structure:
\begin{eqnarray}\label{eq:Jwkpsps}
 \tilde{J}^{\nu}_{B\rightarrow D}
(\vec{{k}}_D^\prime;\vec{{k}}_B=\vec{0})
\es
\frac{\sqrt{k_B^0 k_D^0}}{4
\pi} \int\, \frac{d^3\tilde{k}_{\bar{u}}^\prime}{2 k_b^0}\,
\sqrt{\frac{{\tilde{k}^0_b} \,{\tilde{k}^0_{\bar{u}}}}
{{\tilde{k}^{\prime 0}_c}\, {\tilde{k}^{\prime 0}_{\bar{u}}}}}
 \,\sqrt{\frac{{\tilde{k}^{\prime 0}_{\bar{u}}}+{\tilde{k}^{\prime 0}_c}}
{{{k}^{\prime 0}_c}+{{k}^{\prime 0}_{\bar{u}}}}}\, \sum_{\mu_b,\mu_c^\prime
=\pm \frac{1}{2}}\!\!\!
\bar{u}_{\mu_c^\prime}(\vec{k}_c^\prime)\,\gamma^\nu\,(1-\gamma^5)\,
u_{\mu_b}(\vec{k}_b)\nonumber  \\ &&\times
D_{\mu_b \mu_c^\prime}^{\frac{1}{2}}\left[
{R}_W(\tilde{v}^\prime_{\bar{u}},B_c(v^\prime_{\bar{u}c}))
 {R}_W^{-1}(\tilde{v}^\prime_c,B_c(v^\prime_{\bar{u}c}))\right] \,
\psi^\ast_{D}\,(\vert \vec{\tilde{k}}_{\bar{u}}^\prime\vert)\, \psi_B
 \,(\vert \vec{\tilde{k}}_{\bar{u}}\vert)\, .
\end{eqnarray}
\end{widetext}
Since we  are dealing with velocity states, the momenta in our decay calculation are CM momenta satisfying $\vec{0}=\mathbf{k}_B=\mathbf{k}_{\bar{u}}+\mathbf{k}_b=\mathbf{k}_e+\vec{k}_{\bar{\nu}_e}+\mathbf{k}_D=\mathbf{k}_e+\vec{k}_{\bar{\nu}_e}+\mathbf{k}^{\,\prime}_{\bar{u}}+\mathbf{k}^{\,\prime}_c$. Furthermore, momenta with tilde satisfy $\tilde{\mathbf{k}}_{\bar{u}}+\tilde{\mathbf{k}}_{\bar{b}}=\tilde{\mathbf{k}}_{\bar{u}}^\prime+\tilde{\mathbf{k}}_{\bar{c}}^\prime=0$. Primed momenta with tilde are related to the corresponding momenta without tilde by means of a canonical boost, i.e.
$k^\prime_i=B_c(v^\prime_{\bar{u} c}) \tilde{k}^\prime_i$, $i=\bar{u},c$, where $v^{\prime}_{\bar{u}c}=(k_{\bar{u}}^{\prime}+k_c^{\prime})/\sqrt{(k^{\prime}_{\bar{u}}+k^{\prime}_c)^2}$. Note that due to the decay kinematics $\mathbf{k}_{b}=\tilde{\mathbf{k}}_{b}$ and $\mathbf{k}_{\bar{u}}=\tilde{\mathbf{k}}_{\bar{u}}$. Using these identities we observe that the expression~(\ref{pseudoscalarcurrent}) for the $B\rightarrow D$ transition current which we extracted from the $\nu_e\, B^-\rightarrow e^-\, D^0$ scattering amplitude reduces to Eq.~(\ref{eq:Jwkpsps}), which we got from the $B^-\rightarrow D^0\, e^-\,\bar{\nu}_e$ decay amplitude. Finally, a four-vector decay current $\tilde{J}^\nu_{B\rightarrow D}({p}_D;{p}_B)$ with the covariant structure~(\ref{decompositiontwo}) is obtained from Eq.~(\ref{eq:Jwkpsps}) by going back to physical particle momenta using Eq.~(\ref{eq:Jcovps}).

The form factors calculated from $\tilde{J}^\nu_{B\rightarrow D}({p}_D;{p}_B)$ for time-like momentum transfers are just functions of the four-momentum transfer squared. No dependence on the invariant mass of the whole system is observed as was the case for space-like momentum transfers. This is easily understood, because the invariant mass of the decaying system is just the mass $m_B$ of the decaying meson, which is fixed. Apart of rotations and Lorentz boosts of the whole system there is no freedom left for choosing the decay kinematics such that the four-momentum transfer is $q^2$. But boosting or rotating the system will not change our results for the (time-like) form factors. This is obvious from Eq.~(\ref{eq:Jcovps}), which holds for arbitrary momenta of the decaying meson. In the space-like region, on the other hand, different combinations of Mandelstam $s$ and the scattering angle will give the same momentum transfer $q^2$, which leaves more freedom for choosing the kinematics for form factor calculations. This ambiguity may be considered as a drawback, but it offers also the possibility for extracting form factors in a reference frame, in which physical non-valence contributions, in particular those responsible for $Z$-graphs, and unphysical spurious contributions coming from violation of cluster separability, are minimized. In the following we will give numerical predictions for weak transition form factors in the space- as well as in the time-like region and we will try to estimate the size of the non-valence $Z$-graph contributions and non-physical effects coming from cluster-separability violation.

\section{Numerical studies}\label{sec:numerics}
Once we have an analytical expression for the weak transition current of a heavy-light bound state in terms of constituent currents and wave functions, we can calculate the current numerically and obtain the corresponding form factors, which can be compared with experiments. We will first calculate transition form factors for space-like momentum transfers, as resulting from $\nu_e\,M\rightarrow e^- \, M^\prime$ scattering. Next we will continue these form factors analytically to time-like momentum transfers and compare the analytic continuation with the outcome of the decay calculation $M\rightarrow M^\prime\, e^-\, \bar\nu_e$. The freedom of choosing the kinematics for scattering gives us some control on the influence of non-valence $Z$-graph contributions and cluster-separability-violating effects, whereas the kinematics of the decay calculation is more or less fixed. The comparison of the analytically continued form factors with the result from the decay calculation can thus give us some clues on the size of $Z$-graph contributions and cluster-separability-violating effects for time-like momentum transfers.

\subsection{ Kinematics}\label{sec:kinematics}
Let us first fix the kinematics for $\nu_e\,M\rightarrow e^- \, M^\prime$ scattering. Since our model current $ \tilde{J}^{\nu}_{M\rightarrow M^\prime}$ is a four-vector with the structure given by Eq.~(\ref{eq:Jcovdec}), form factors are neither affected by rotations or boosts. We can thus consider $\nu_e\,M\rightarrow e^- \, M^\prime$ scattering in the CM frame, $\sum_{i=1}^n \vec p_i = 0$, and also fix the scattering plane to the $(1,3)$-plane. Unlike the physical form factors, our model form factors exhibit a Mandelstam $s$ dependence which can be reinterpreted as a frame dependence of the $W\,M\rightarrow M^\prime$ subprocess~\cite{Biernat:2009my}. In order to obtain an estimate of this frame dependence we will consider two extreme cases. In the first case $s$ has the minimal value for reaching a particular value of $q^2$. This corresponds to the Breit frame. In the second case we will consider the limit $s\rightarrow\infty$, corresponding to the IMF. For CM scattering in the (1,3)-plane our kinematics is defined by:

\begin{eqnarray}\label{eq:cmkinem}
\ul p_M
&=&
\left(
\begin{array}{c}
\sqrt{m_M^2 + p^2 }\\
p \\
0 \\
0
\end{array}
\right) , \quad\!\!
p_{\nu_e}
\ = \
\left(
\begin{array}{c}
\sqrt{m_{\nu_e}^2 + p^2 }\\
- p\\
0 \\
0
\end{array}
\right)  , \nonumber\\
\ul p_{M^\prime}
&=&
\left(
\begin{array}{c}
\sqrt{m_{M^\prime}^2 + p'^2 } \\
p^\prime \,\cos\theta_{\mathrm{CM}}\\
0 \\
p^\prime \,\sin\theta_{\mathrm{CM}}
\end{array}
\right)  , \quad\!\!
p_{e}
\ = \
\left(
\begin{array}{c}
\sqrt{m_{e}^2 + p'^2 }\\
-p^\prime \,\cos\theta_{\mathrm{CM}}\\
0 \\
-p^\prime \,\sin\theta_{\mathrm{CM}}
\end{array}
\right) , \nonumber\\
\end{eqnarray}
where $\theta_\mathrm{CM}$ is the CM scattering angle. The kinematics can also be expressed in terms of two Lorentz invariants, namely Mandelstam $s=(p_M+p_{\nu_e})^2=(p_{M'}+p_e)^2$, the invariant mass squared of the whole system, and $q^2=(p_M-p_{M'})^2=(p_e-p_{\nu_e})^2$ (also termed as Mandelstam $t$), the four-momentum transfer squared. Neglecting the lepton masses, one has
\begin{eqnarray}\label{eq:kinemmandel}
p _M&=& {(s - m_{M}^2)  \over 2\sqrt{s}} \, , \quad
p_{M^\prime} = {(s - m_{M'}^2)  \over 2\sqrt{s}} \, , \nonumber\\ && \hspace{-1.0cm}\cos\theta_\mathrm{CM}=-\frac{Q^2}{2 p p'}+1\, .
\end{eqnarray}
with space-like $q^2 = q^\mu q_\mu =: -Q^2<0$. This guarantees that $-1\leq \cos\theta_\mathrm{CM}\leq 1$.

\subsubsection{Infinite-momentum frame}
The infinite-momentum frame is defined by the limit $s\to\infty$, $Q^2=-q^2=\mathrm{const}$. This limit implies for the CM kinematics given above that the components of the current, Eq.~(\ref{decompositiontwo}), take on the simple form:
\begin{eqnarray}
 J^0(p_{M'},p_{M}) &\rightarrow&  \sqrt{s} \, F_1\,, \nonumber\\ J^1(p_{M'},p_{M})  &\rightarrow& \sqrt{s} \, F_1  \,,  \\
 J^2(p_{M'},p_{M}) &\rightarrow& 0\,,\hspace{0.9cm}\nonumber\\ J^3(p_{M'},p_{M}) &\rightarrow& \frac{(m_{M}^2-m_{M'}^2) \, F_0- (m_{M}^2-m_{M'}^2+Q^2)\, F_1}{Q} \ .\nonumber
\end{eqnarray}
Taking $J^0$ and $J^3$ as linear independent components,  the form factors are then given by:
 \begin{eqnarray}
  F_1&=& \frac{J^0(p_{M'},p_{M})}{\sqrt{s}}\, , \\ F_0&=& \frac{\sqrt{s}\,Q\,J^3(p_{M'},p_{M})+(m_{M}^2-m_{M'}^2-Q^2)\,J^0(p_{M'},p_{M})}{(m_{M}^2-m_{M'}^2)\,\sqrt{s}}.\nonumber
 \end{eqnarray}

\subsubsection{Breit frame}
What we call \lq\lq Breit frame\rq\rq\  corresponds to backward scattering, i.e. $\theta_{\mathrm{CM}}=\pi$. For $M=M'$ it reduces to the usual Breit frame, which is defined by zero energy transfer. In the BF Mandelstam $s$ and $Q^2$ are related by means of Eq.~(\ref{eq:kinemmandel}) and we can express Mandelstam $s$ as function of $Q^2$:
\begin{eqnarray}\label{eq:sbreit}
s&=&\frac{1}{2}\bigg( m_{M}^2+m_{M'}^2+Q^2 \\&&\left. +\sqrt{(m_{M}^2+m_{M'}^2+Q^2)^2-4 m_{M}^2 m_{M'}^2} \right) \ .\nonumber
\end{eqnarray}
This is the minimum value of Mandelstam $s$, necessary to achieve a momentum transfer $Q^2$. The only non-vanishing current components for this kinematics are $J^0$ and $J^1$, from which $F_0$ and $F_1$ can be determined uniquely using Eq.~(\ref{eq:Jcovdec}).

\subsection{Frame dependence of space-like transition form factors}\label{sec:spacelike}
\label{framedep}
\begin{figure}[t!]
\includegraphics[scale=0.41]{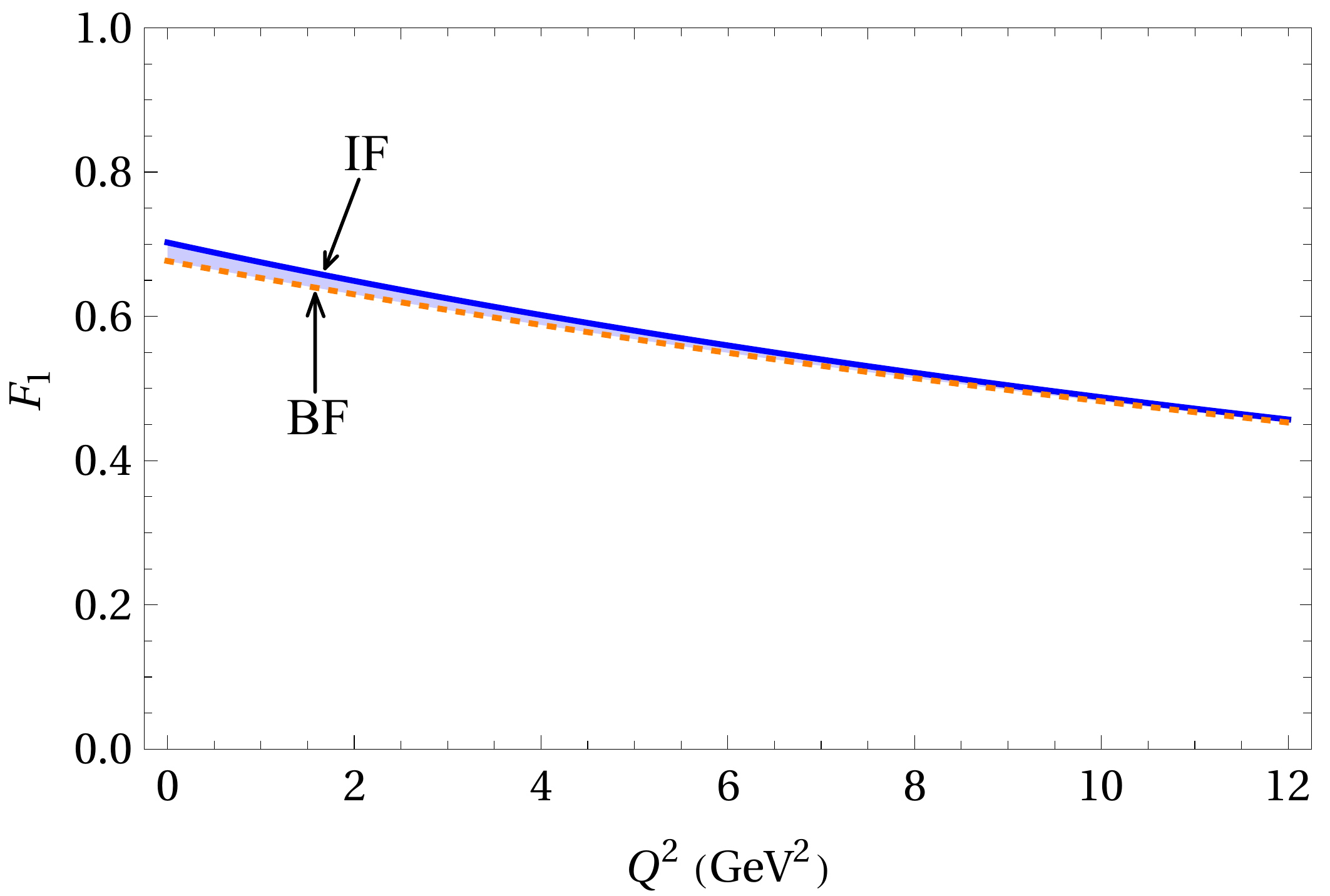}
\includegraphics[scale=0.7]{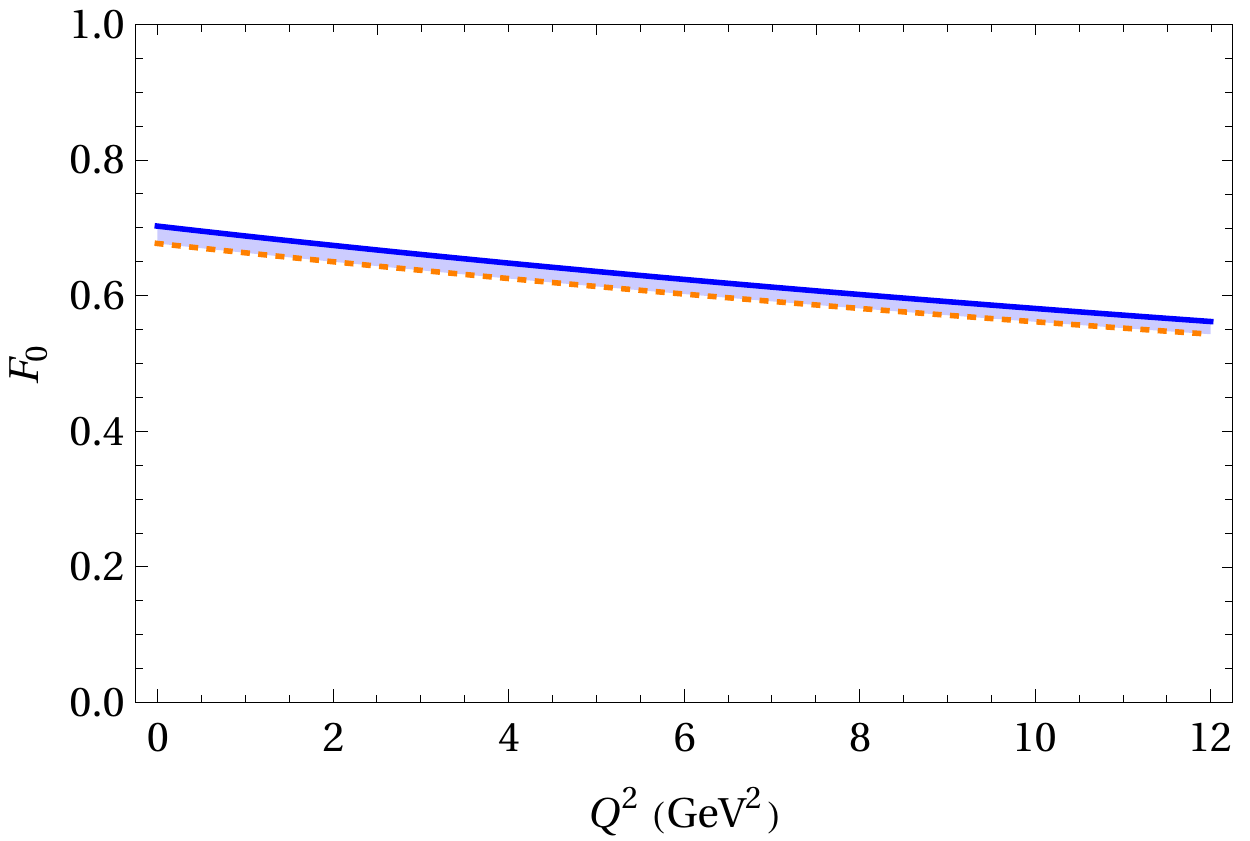}
\caption{ (Color online) Space-like form factors for the $B^- \to D^0$ transition calculated in the IMF (solid line) and in the BF (dashed line). The shaded area indicates the frame dependence. }
\label{BtoDframes}
\end{figure}

\begin{figure}[t!]
\includegraphics[scale=0.7]{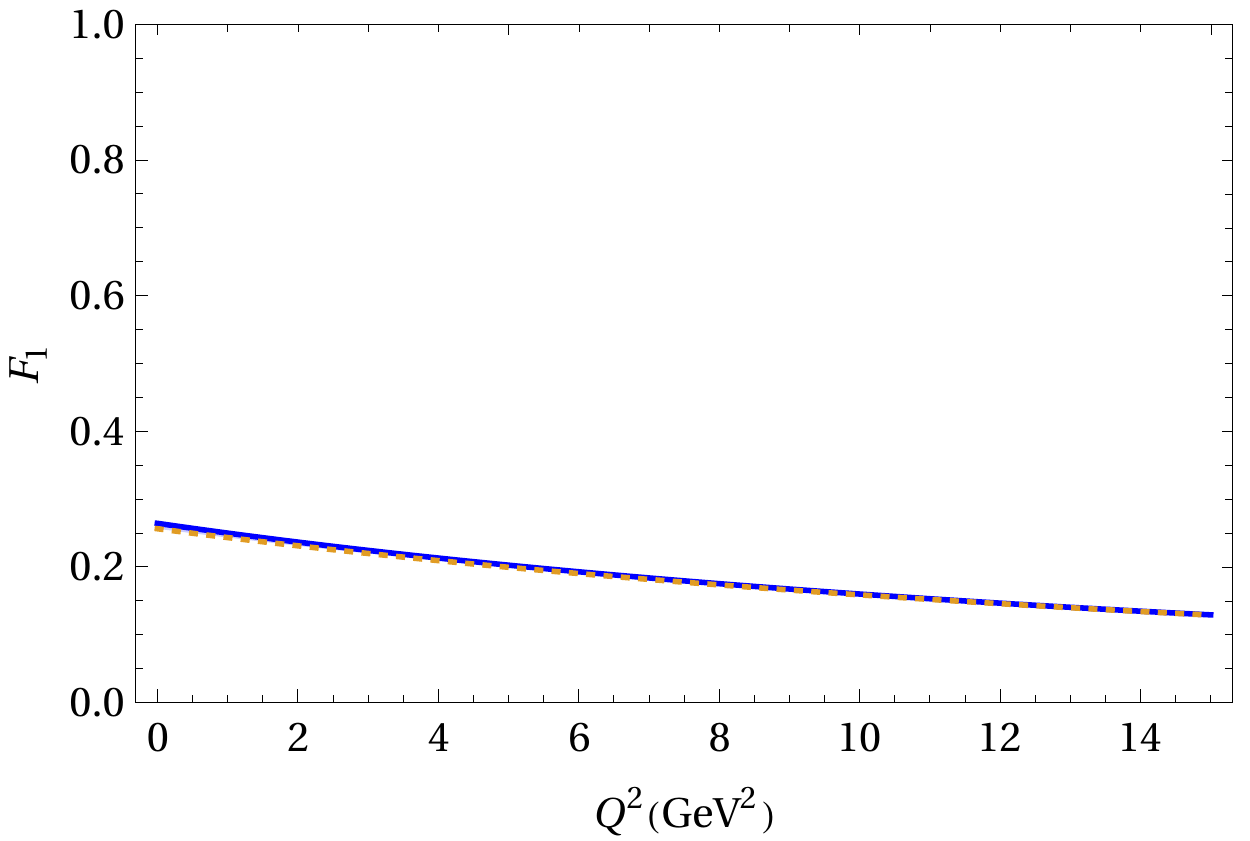}
\includegraphics[scale=0.7]{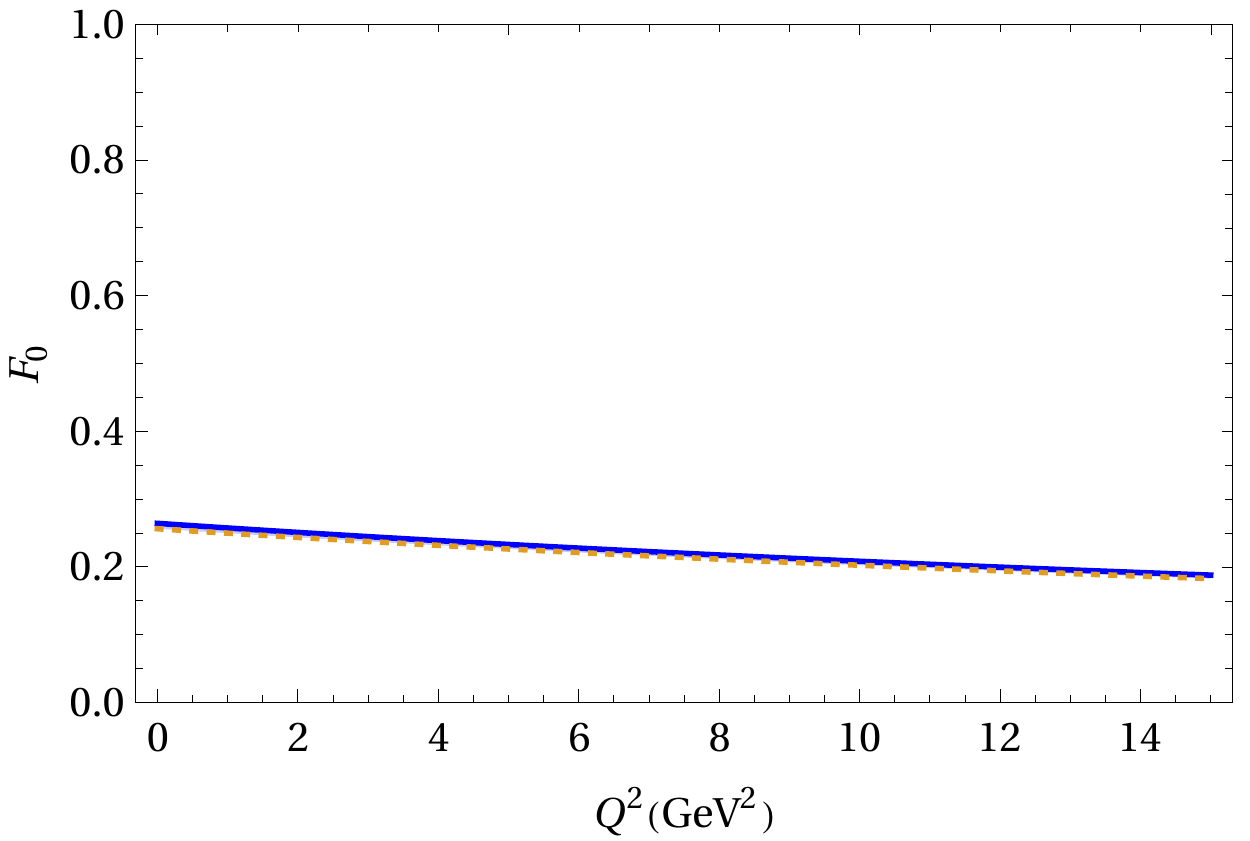}
\caption{(Color online) Space-like form factors for the $B^-\to \pi^0 $ transition calculated in the IMF (solid line) and in the BF (dashed line). The shaded area indicates the frame dependence.}
\label{BtoPiframes}
\end{figure}

\begin{figure}[t!]
\includegraphics[scale=0.7]{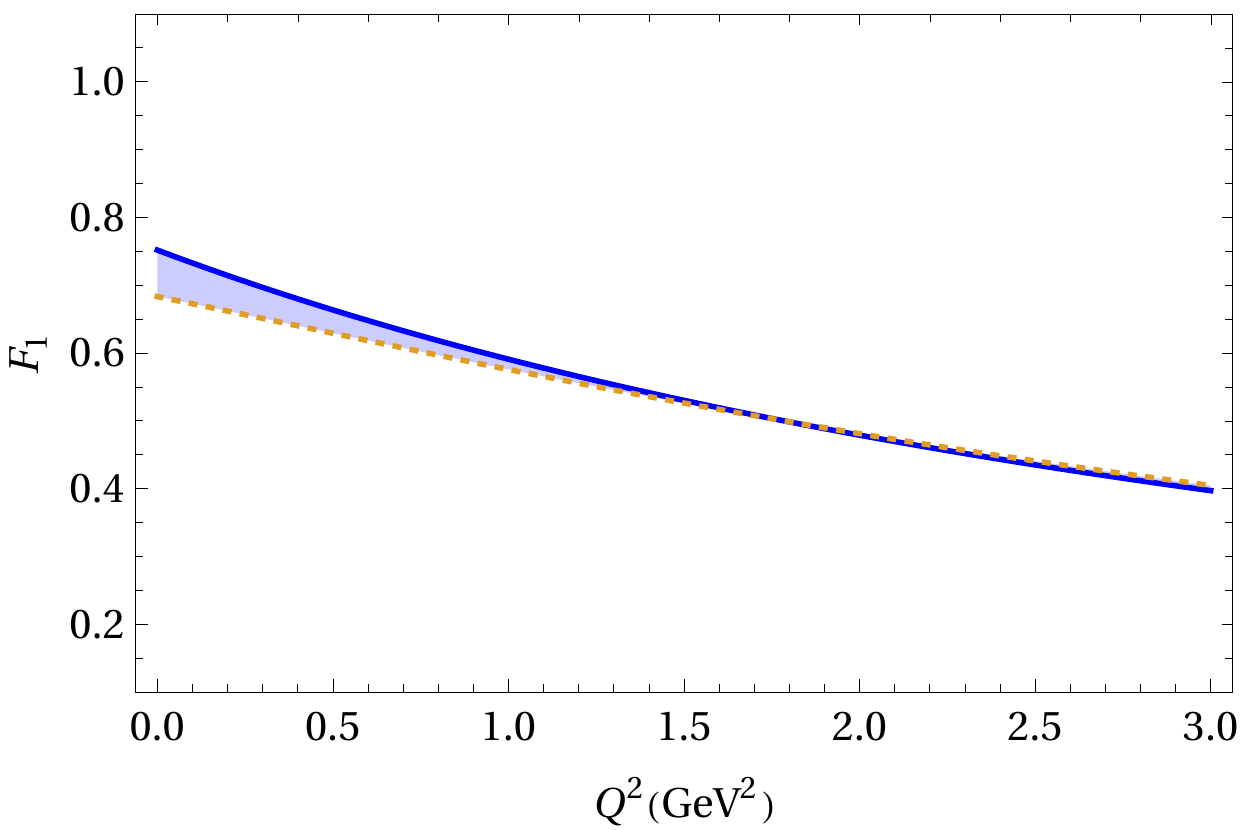}
\includegraphics[scale=0.7]{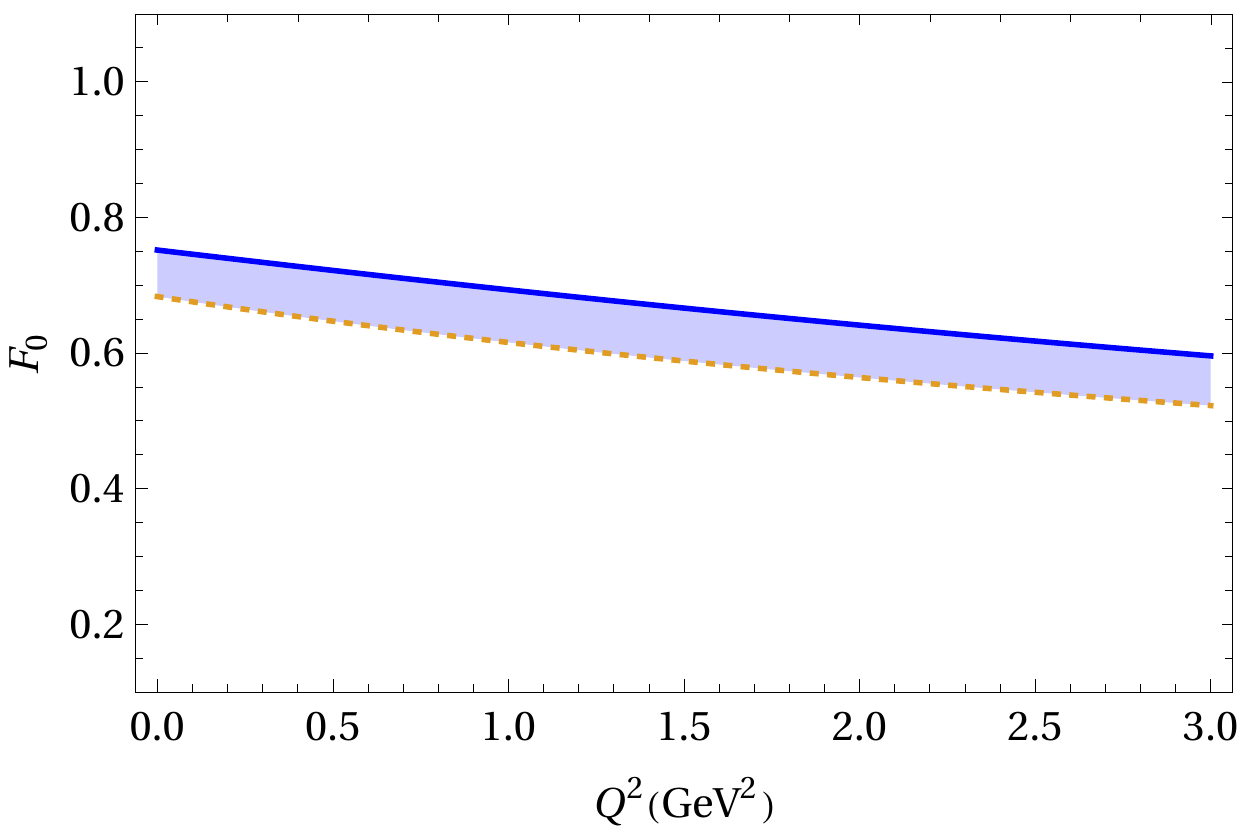}
\caption{ (Color online) Space-like form factors for the $D^- \to K^0 $ transition calculated in the IMF (solid line) and in the BF (dashed line). The shaded area indicates the frame dependence.}
\label{DtoKframes}
\end{figure}

\begin{figure}[t!]
\includegraphics[scale=0.7]{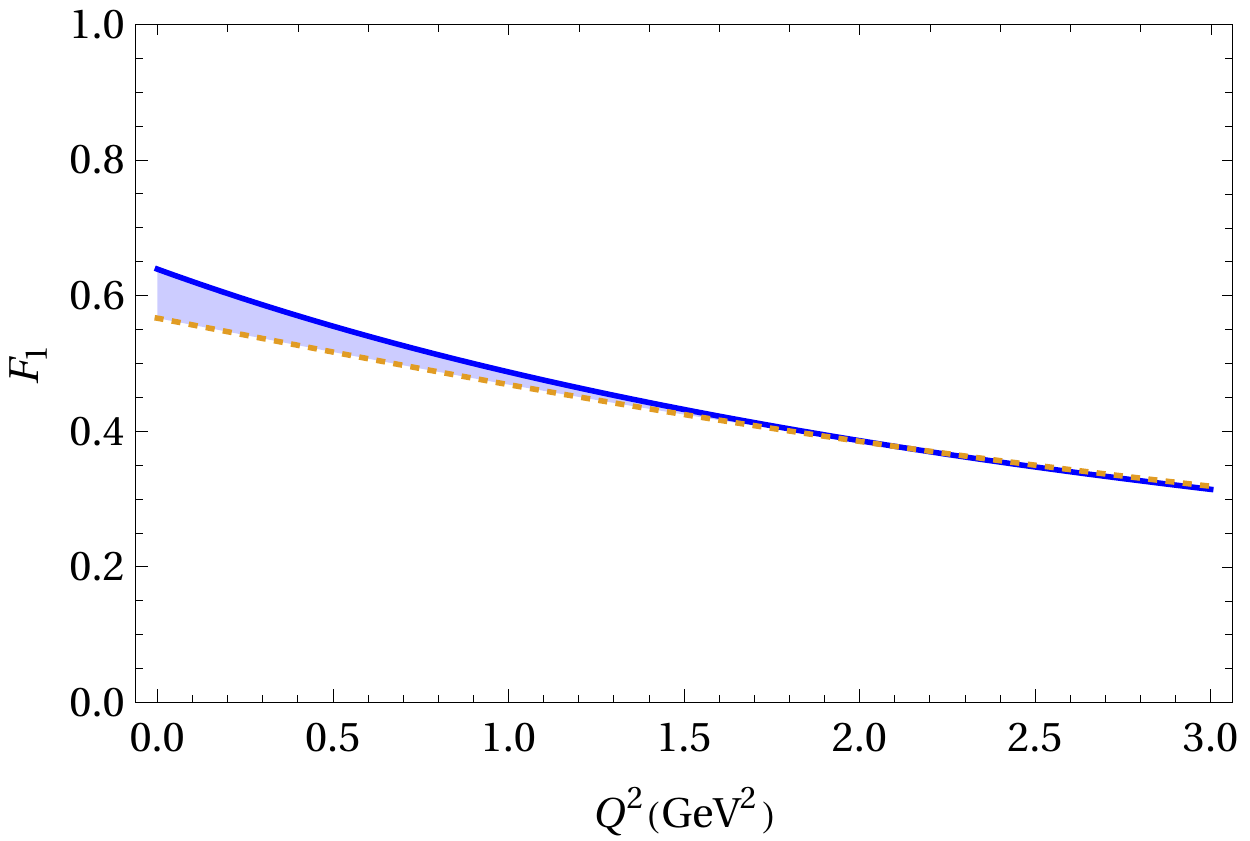}
\includegraphics[scale=0.7]{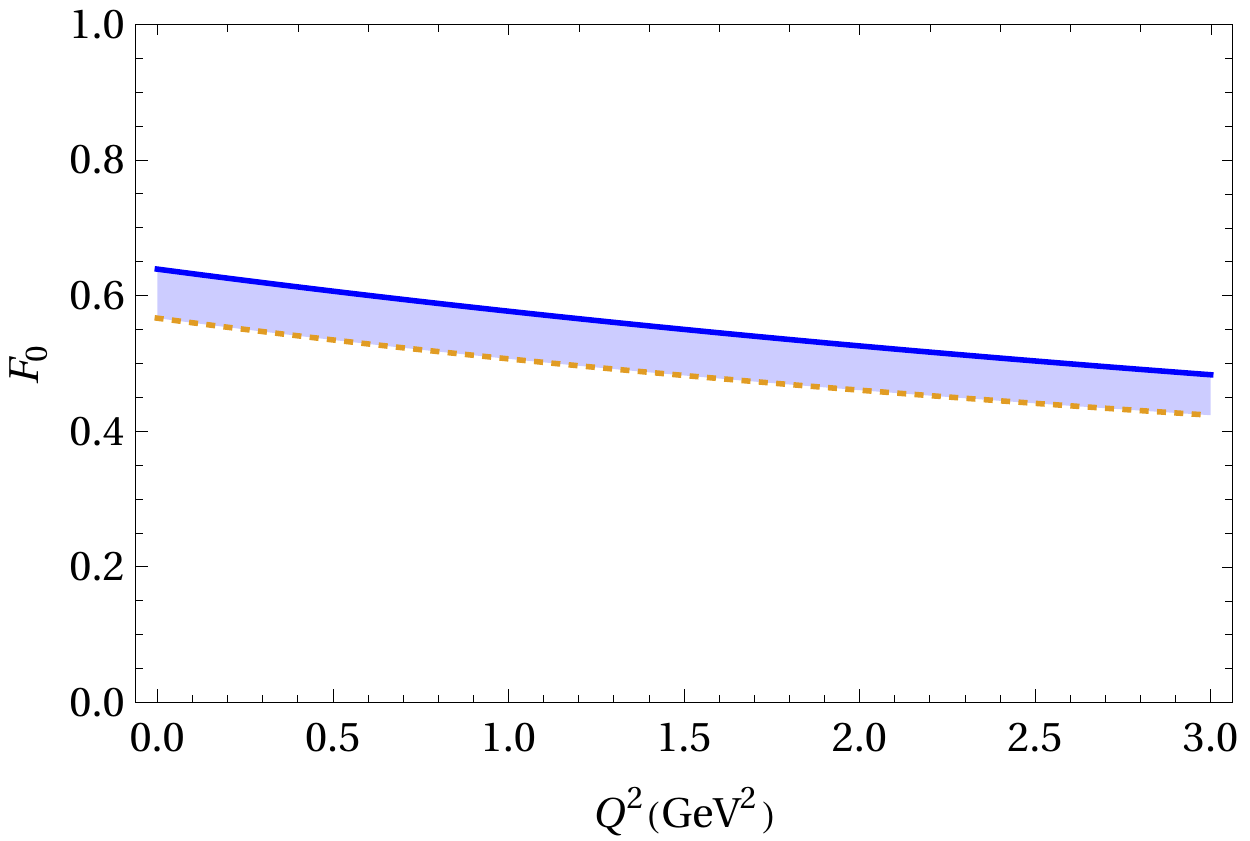}
\caption{ (Color online) Space-like form factors for the $D^- \to \pi^0$ transition calculated in the IMF (solid line) and in the BF (dashed line). The shaded area indicates the frame dependence.}
\label{DtoPiframes}
\end{figure}

In order to estimate the frame dependence of our form factors, we now consider the two extreme cases of minimal $s$, i.e. the Breit frame, and $s\rightarrow\infty$, the infinite-momentum frame. Calculating the current (\ref{pseudoscalarcurrent}) in the respective frames with the wave functions and model parameters given in App.~\ref{app:wf}  and applying the covariant decomposition (\ref{eq:Jcovdec}) gives the form factor results shown in Figs.~\ref{BtoDframes} to \ref{DtoPiframes}. In addition to the $B^-\rightarrow D^0$ transition we have also considered the $B^-\rightarrow\pi^0$, the $D^-\rightarrow K^0$ and the $D^-\rightarrow \pi^0$ transitions to get a feeling on how the frame dependence is affected by the heavy-quark flavors and the masses of the initial- and final-state mesons.

What one observes are rather small differences between the IMF and the BF predictions for the $B\rightarrow D$ and the $B\rightarrow \pi$ transition form factors. The discrepancies between the two frames become a little bit more pronounced for the $D\rightarrow K$ and the $D\rightarrow \pi$ transition form factors. This has a simple explanation: As can be seen from Eq.~(\ref{eq:sbreit}), for fixed $Q^2$ the value of Mandelstam $s$ in the BF is larger for larger mass of the decaying meson and thus closer to the infinite $s$ value in the IMF. With increasing $Q^2$ also $s$ has to increase in the BF and thus the BF result tends to the IMF result for large $Q^2$. Another observation is that, for fixed initial-state meson $M$, the discrepancies become smaller for smaller mass of the final-state meson $M'$. This indicates that effects due to wrong cluster properties decrease, if $p'$, the size of the three-momentum of the final-state meson (see Eq.~(\ref{eq:cmkinem})), increases. For large enough $p'$ bound-state effects, and thus also effects coming from wrong cluster properties, become negligible. Obviously $F_0$ exhibits a stronger frame dependence than $F_1$. This is also not too surprising, since $F_0=F_1+F^-$ (see Eqs.~(\ref{decompositionone}) and (\ref{eq:f0f1})).  In addition, there seems to be a stronger frame dependence in the small form factor $F^-$ than in the large form factor $F^1=F^+$.

These observations are in line with foregoing work~\cite{Biernat:2009my,Biernat:2010tp,GomezRocha:2012zd,Biernat:2014dea} and suggest that the frame dependence of the form factors and hence the influence of the $Z$-graph and unwanted effects coming from wrong cluster properties of our approach tend to diminish with increasing value of Mandelstam $s$. It occurs thus to be preferable to use IMF kinematics for the extraction of the transition form factors. A further advantage of the IMF is that a valence-quark approach catches more of the physics in the IMF than in any other reference frame.  The comparison of the IMF with the BF results may thus also give us a hint on the size of $Z$-graph contributions in reference frames different from the IMF. Since physical form factors should not depend on the chosen reference frame, we consider the results obtained with IMF kinematics as the more complete one.

\subsection{Time-like transition form factors -- analytic continuation of BF results}\label{sec:BFanalytic}
Lepton-meson scattering amplitudes (and hence also transition currents and form factors) are meromorphic functions of Mandelstam  $s$ and $t$ ($=q^2=q^\mu q_\mu$). It is therefore possible to continue them analytically from $q^2\le 0$ to $q^2\ge 0$. This amounts to replace $Q$ by $i\,Q$ in the analytical expressions for the transition current and the form factors, i.e. Eqs.~(\ref{pseudoscalarcurrent}) and $(\ref{eq:Jcovdec})$, respectively. Due to the reasons just mentioned, it seems to be preferable to perform the analytic continuation with IMF kinematics as starting point.

\begin{figure}
\includegraphics[scale=0.7]{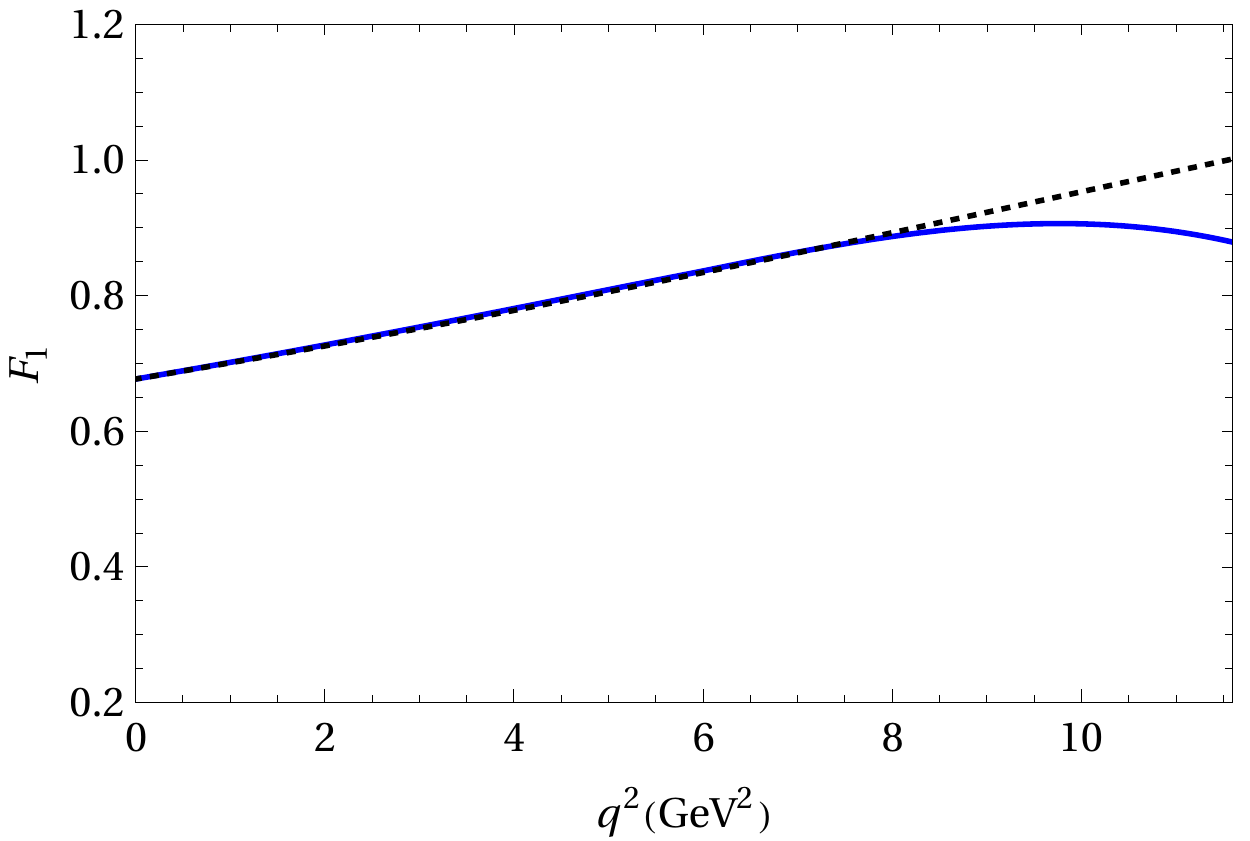}
\includegraphics[scale=0.7]{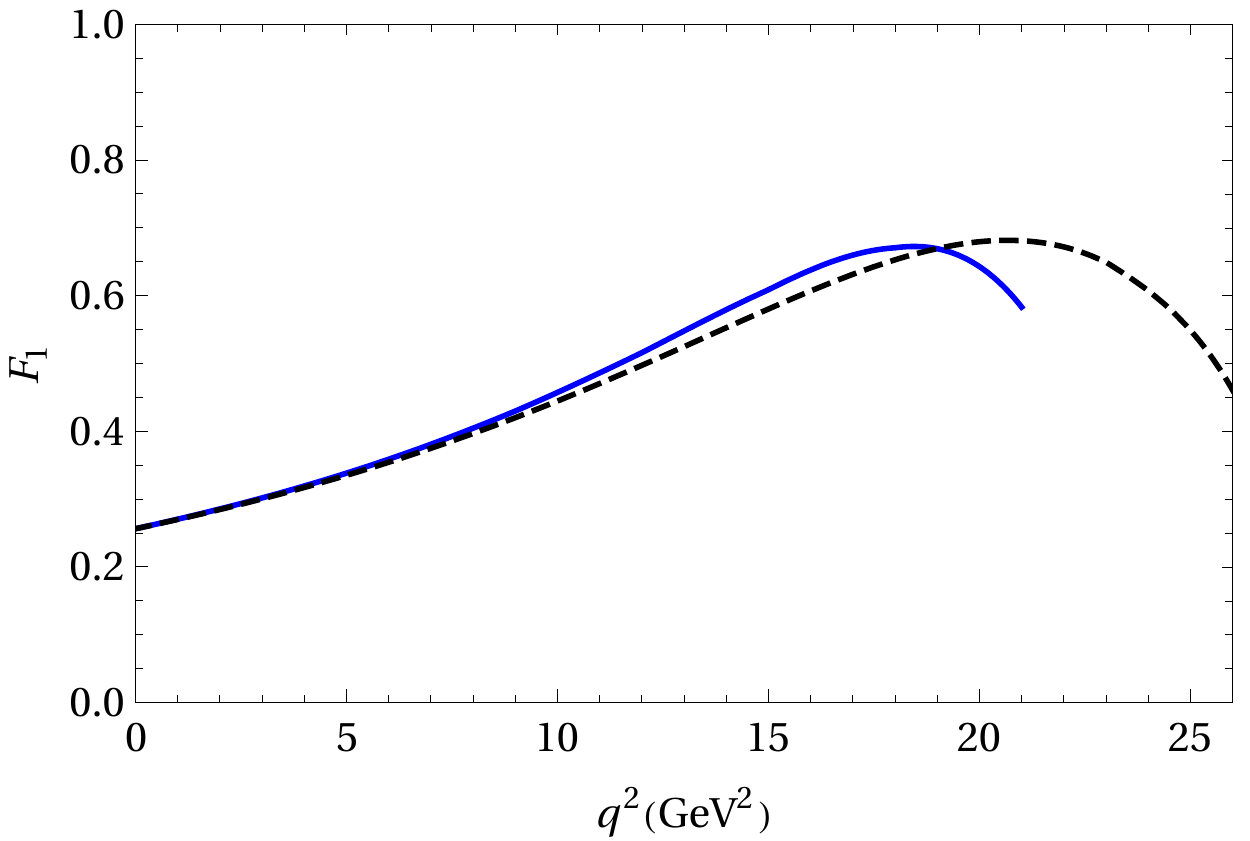}
\caption{ (Color online) The form factor $F_1$ for the $B\rightarrow D$ (upper panel) and $B\rightarrow\pi$  (lower panel) transition for time-like momentum transfers $q^2\geq 0$. The solid line corresponds to the analytic continuation of the BF results, the dashed line is the outcome of the direct decay calculation  (cf. Eq.~(\ref{eq:Jwkpsps})). }
\label{Breit-AC}
\end{figure}

Nevertheless, it is also instructive to have a look on analytic continuation starting from BF kinematics. This frame is energetically closest to the decay kinematics, for which the invariant mass squared of the decaying system is $s=m_B^2=\text{const}$. Analytic continuation of the BF results is thus expected to resemble the outcome of the direct decay calculation and, if this is the case, give us some confidence in the reliability of the analytic continuation procedure. A comparison of the analytic continuation $Q\to i\,Q$ of the BF results with the direct decay calculation, as outlined in Sec.~\ref{sec:decay},  is shown in Figure~\ref{Breit-AC} for the $B\to D$ and $B\to \pi$ transitions in the physically allowed range $0\leq q^2\leq q^2_{\mathrm{max}}=(m_M-m_{M'})^2$. We observe that the predictions from analytic continuation and the direct decay calculation agree at $q^2=0$ for both transitions. The agreement extends approximately up to $q^2\approx 8 \text{GeV}^2$, where the differences start to increase with increasing $q^2$ and become significant towards the zero-recoil point $q_{\mathrm{max}}^2=(m_B - m_{D(\pi)})^2$. This behavior is easily understood: The direct decay calculation is done at fixed $s=m_B^2$. In the analytically continued BF calculation, however, $s$ decreases with increasing $Q^2$ (see Eq.~(\ref{eq:sbreit})) from $s=m_B^2$ to $s=m_B m_D$ (or $s=m_B m_\pi$). This means that the form factors obtained by analytic continuation and by the direct decay calculation are calculated for the same $q^2$ value, but for different values of Mandelstam $s$. The physical form factors should, of course, be independent of $s$ and thus independent of the frame in which the $WMM'$ vertex is considered, but as already mentioned before, our model results exhibit an $s$-dependence which can have its origing in a missing $Z$-graph contribution and in wrong cluster properties. With increasing $q^2$ the BF kinematics differs more and more from the decay kinematics with $s=m_B^2$ fixed and, consequently, the $s$-dependence leads then to the observed discrepancies between the analytic continuation of the BF results and the direct decay calculation.

\subsection{Time-like transition form factors -- analytic continuation of IMF results}\label{sec:IMFanalytic}

\begin{figure}[b!]
\includegraphics[scale=0.38]{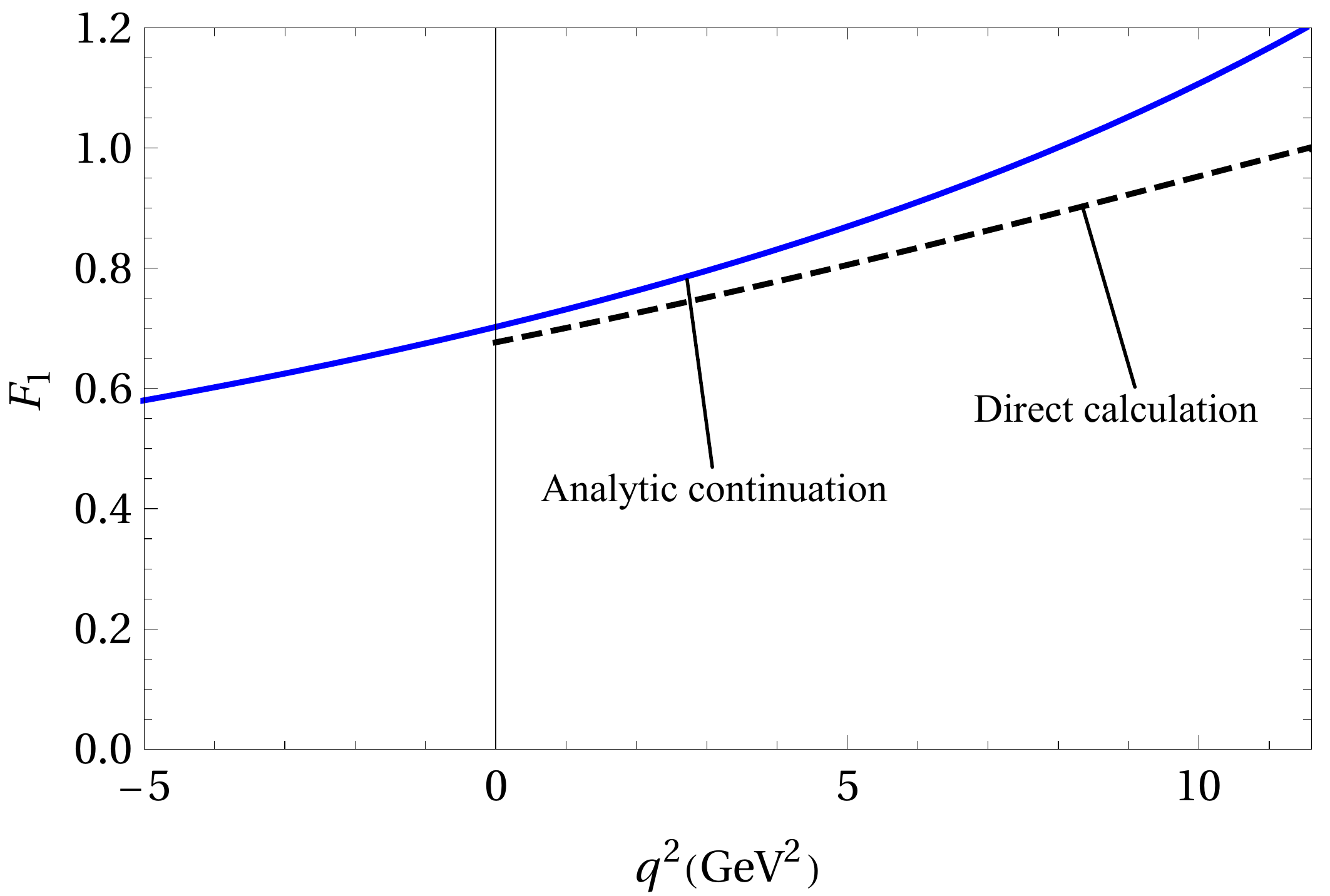}
\includegraphics[scale=0.65]{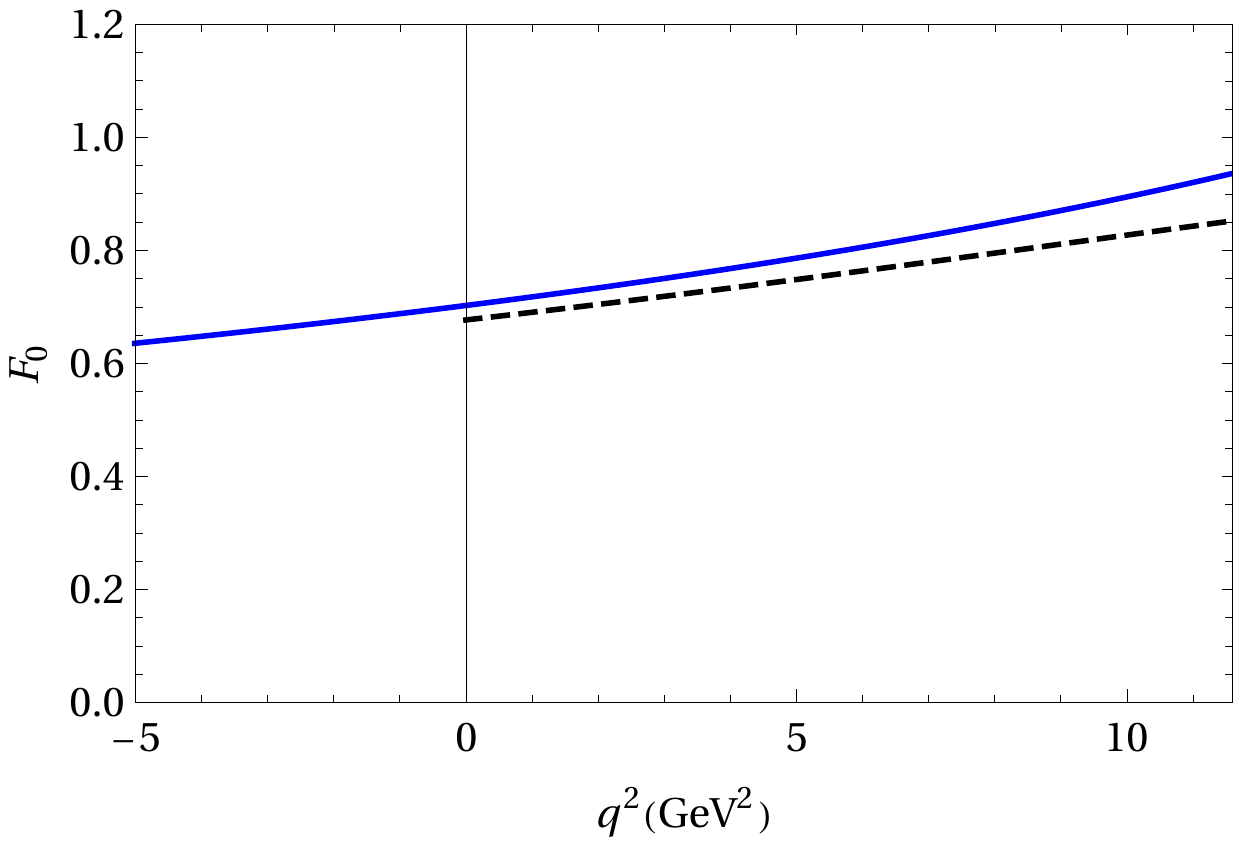}
\caption{ (Color online) The form factors $F_1$ and $F_0$ for the $B\rightarrow D$ transition for time-like momentum transfers $q^2\geq 0$. The solid line corresponds to the analytic continuation of the IMF results, the dashed line is the outcome of the direct decay calculation  (cf. Eq.~(\ref{eq:Jwkpsps})).}
\label{ZBtoD}
\end{figure}

\begin{figure}[b!]
\includegraphics[scale=0.65]{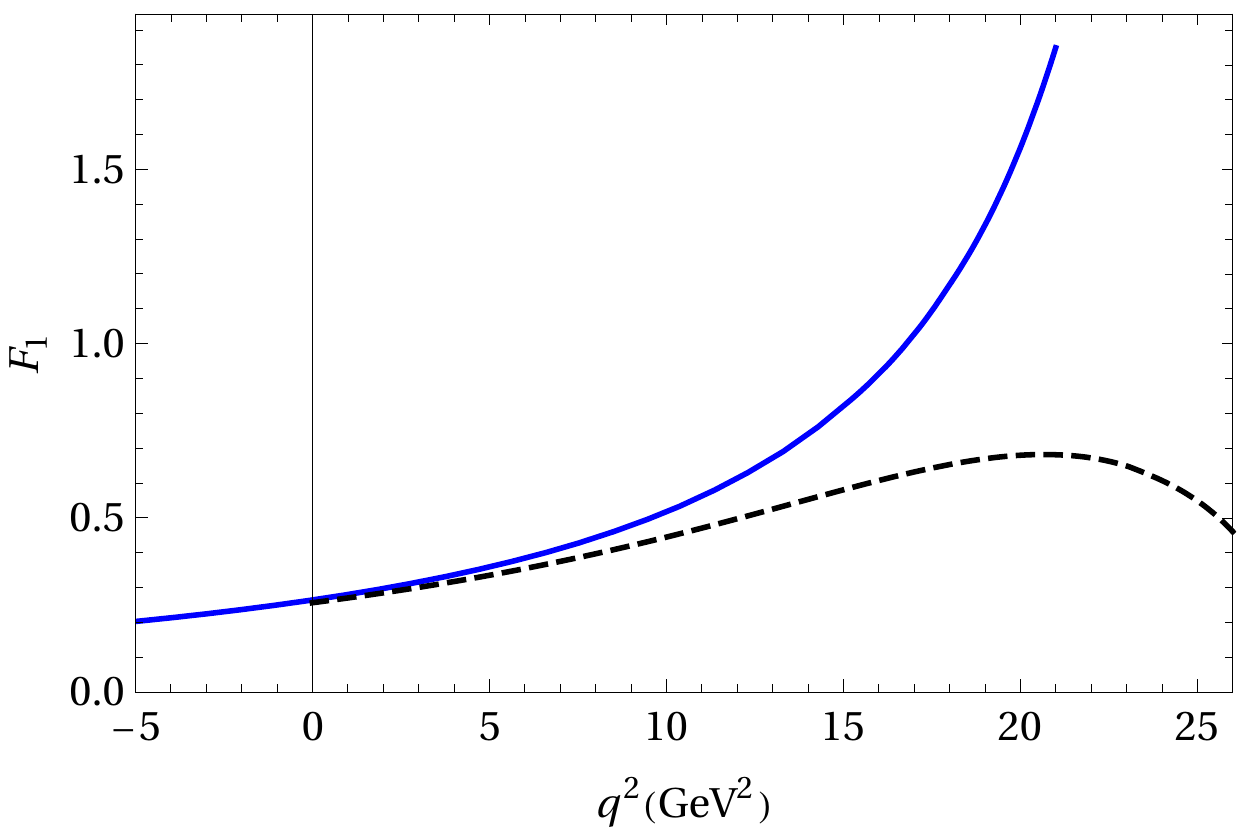}
\includegraphics[scale=0.65]{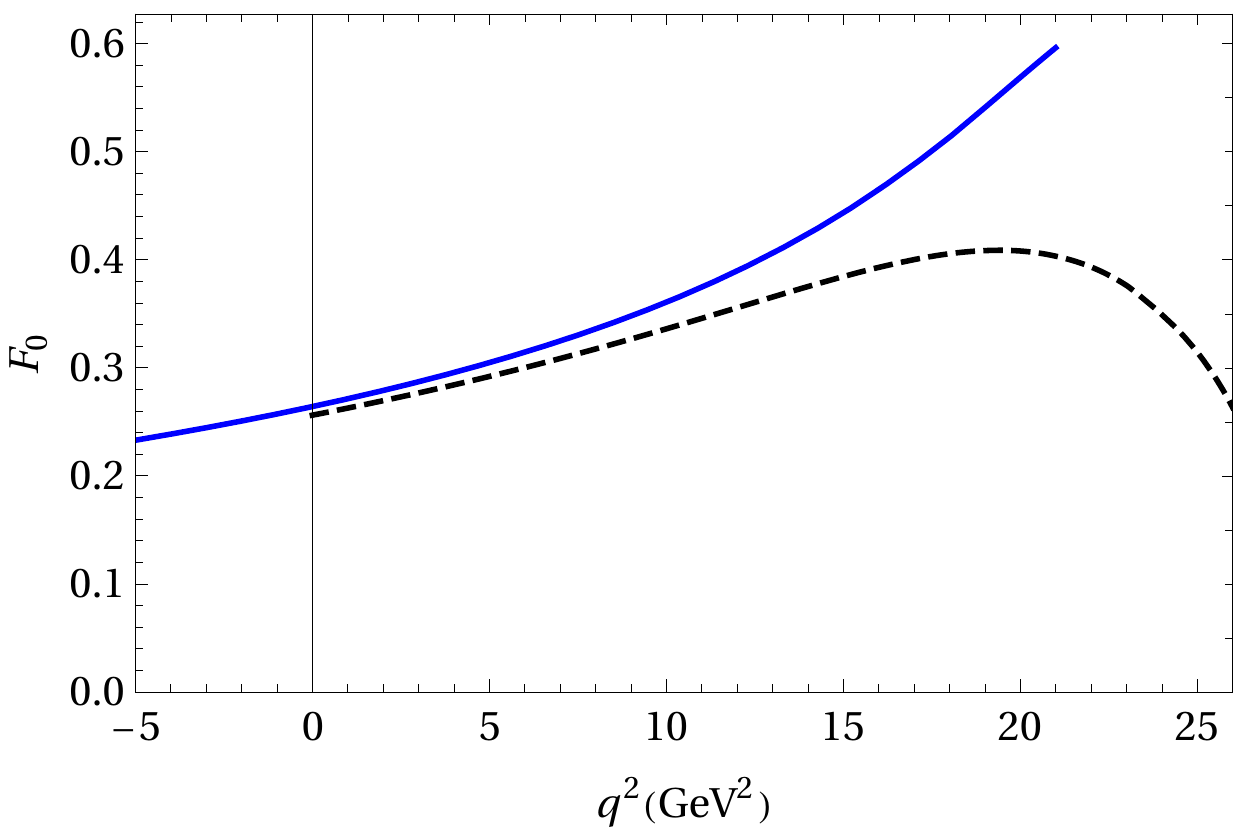}
\caption{ (Color online) Same as in Figure~\ref{ZBtoD} but for the $B\to \pi$ transition.  }
\label{ZBtoPi}
\end{figure}

\begin{figure}[b!]
\includegraphics[scale=0.65]{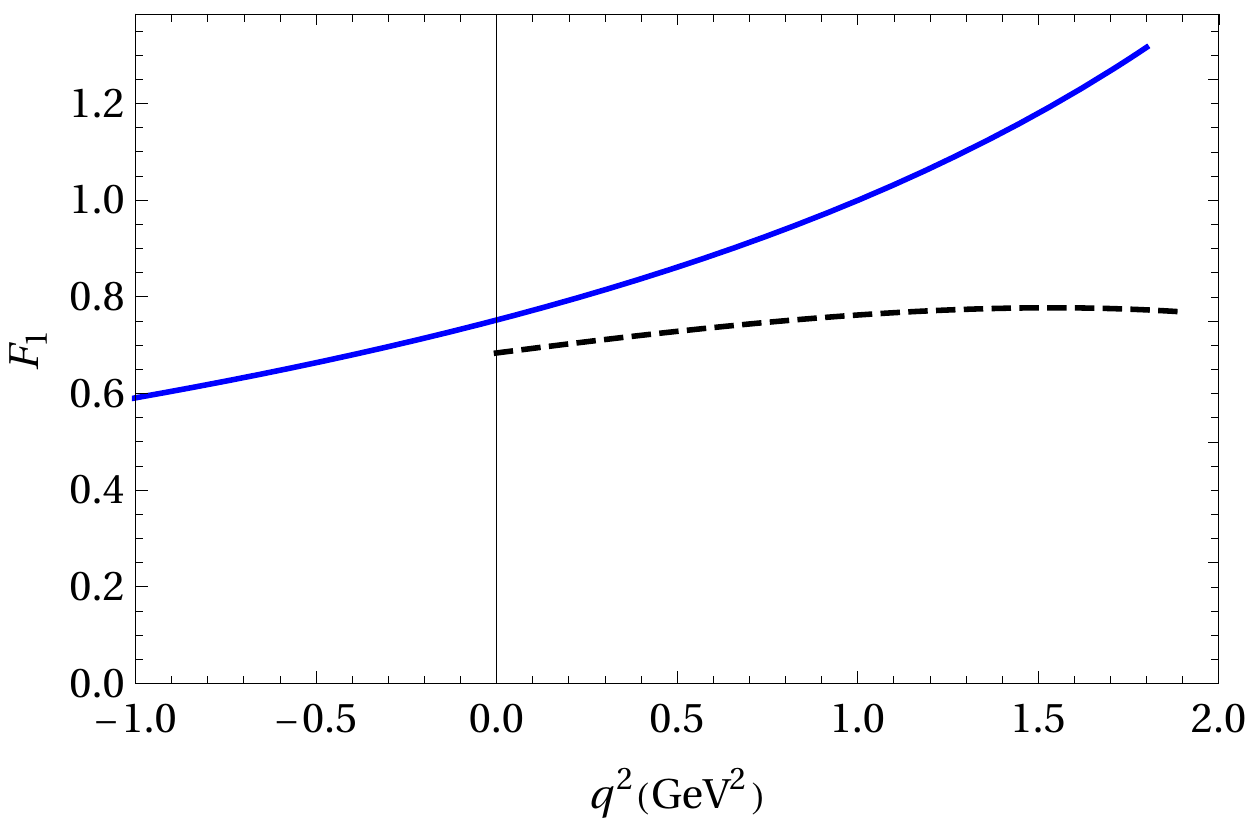}
\includegraphics[scale=0.65]{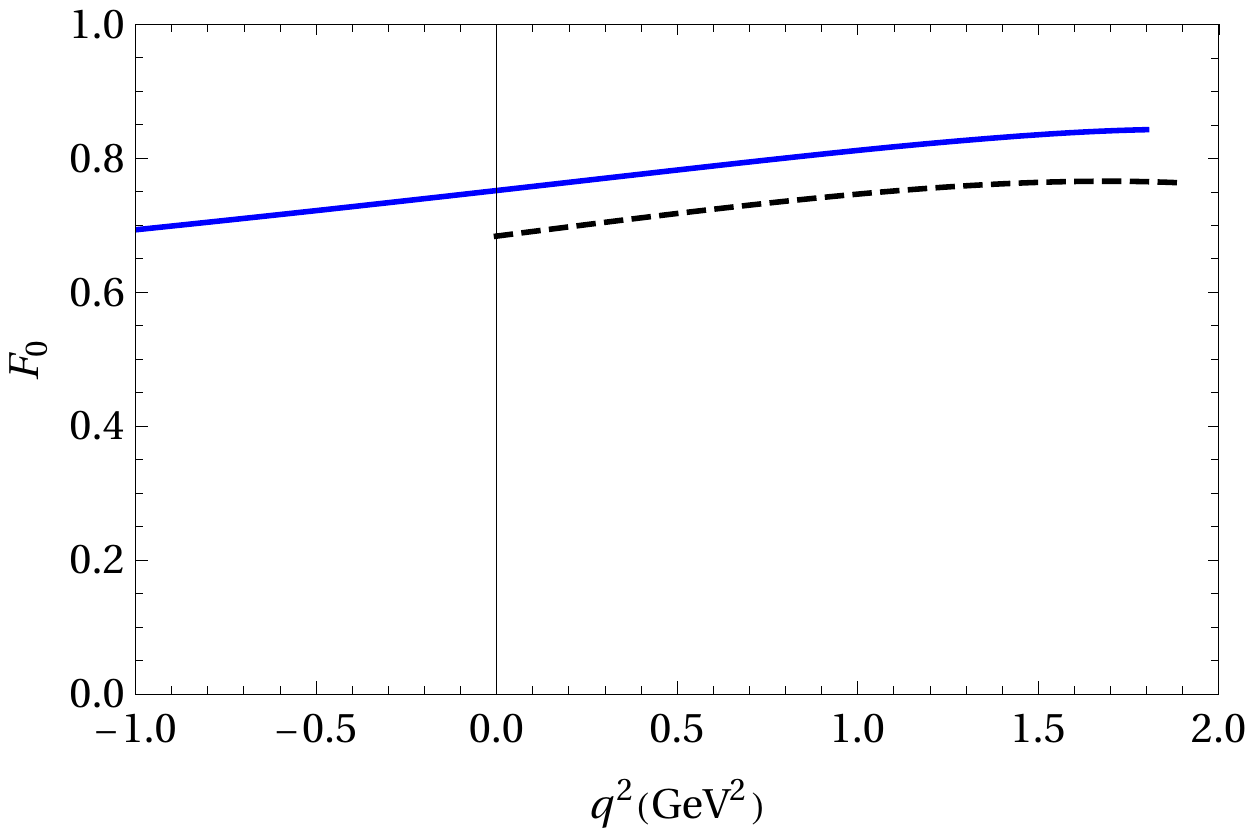}
\caption{ (Color online) Same as in Figure~\ref{ZBtoD} but for $D\to K$ transition.  }
\label{ZDtoK}
\end{figure}

\begin{figure}
\includegraphics[scale=0.65]{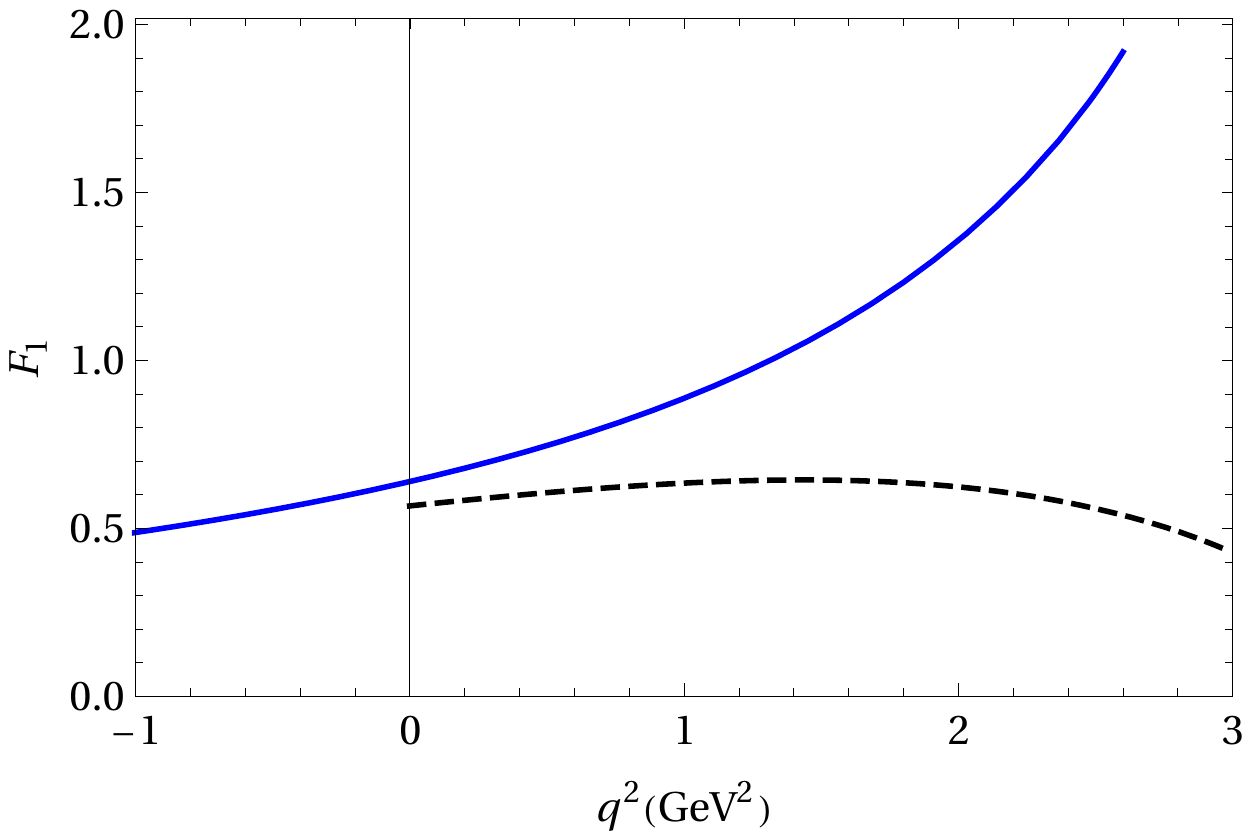}
\includegraphics[scale=0.65]{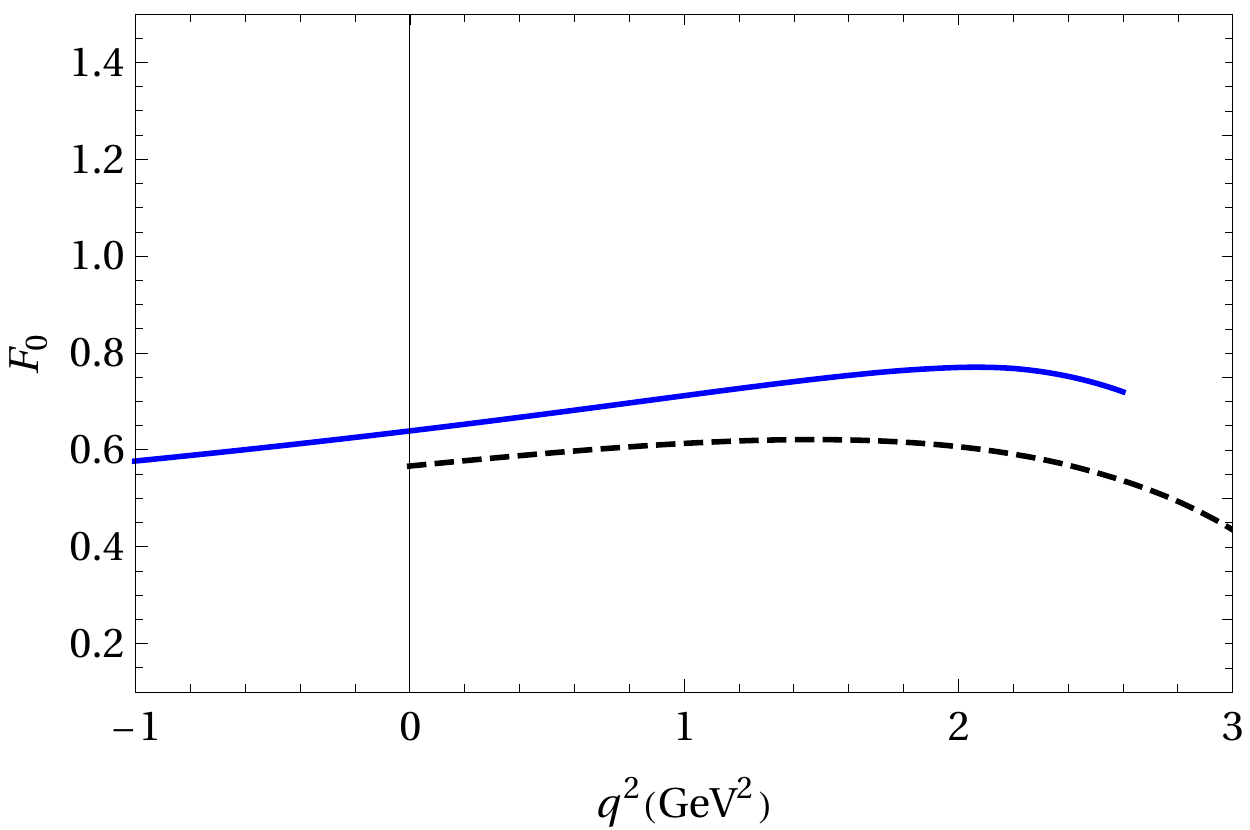}
\caption{ (Color online)  Same as in Figure~\ref{ZBtoD} but for $D\to \pi$ transition. }
\label{ZDtoPi}
\end{figure}

Since cluster-separability-violating effects tend to be minimized in the IMF (i.e. for $s\to\infty$) and non-valence $Z$-graph contributions are suppressed, it seems to be preferable to start with form factor expressions calculated in the IMF and apply analytic continuation to those expressions. In the IMF, furthermore, $s$ is fixed so that the analytic continuation procedure is less delicate. The comparison of the analytically continued IMF results with the direct decay calculation, performed in the rest frame of the decaying particle ($s=m_B^2$), will give us then an estimate of the size of cluster-separability-violating effects and the size of non-valence $Z$-graph contributions in the rest frame of the decaying particle.

In Figs.~\ref{ZBtoD}--\ref{ZDtoPi} the analytic continuation of the IMF results for $F_0$ and $F_1$ is compared with the direct decay calculation for various heavy-light transitions. What one observes is, that the differences between the analytic continuation and the direct decay calculation increase with increasing $q^2$ and are, in general, smaller for $F_0$ than for $F_1$. The differences become also larger, when the mass of the outgoing or decaying meson becomes smaller. For the $B\rightarrow\pi$ and $D\rightarrow\pi$ transitions a strong increase of the IMF result near the zero-recoil point $q_{\mathrm{max}}^2=(m_{B(D)}-m_{\pi})^2$ can be observed for $F_1$ which resembles a pole-like behavior. One can find several constituent-quark-model calculations for the transition-form factors considered in this paper in the literature. The outcome of these calculations resembles very strongly our IMF results. In order to show the variations between different, more recent model predictions, we have summarized numerical form factor values at zero momentum transfer in Tab.~\ref{comparisons}.

\begin{widetext}
\begin{center}
\begin{table}[t!]
\caption{Form factors at zero momentum transfer, $F_1 (0)=F_0(0)$, for $B\rightarrow D$, $\pi$ and $D\rightarrow K$, $\pi$ transitions, as predicted by our model, in comparison with other constituent-quark-model results. }
\begin{tabular}{ccccccccc}
\hline
\hline
Transition              & \textbf{This } &  Zhang \textit{et al.} & Choi & Faustov \textit{et al.} & Verma & Cheng \textit{et al.} & Ivanov \textit{et al.} & Wu \textit{et al.} \\
              & \textbf{work} &  \cite{Zhang:2020dla} & \cite{Choi:2021qza,Choi:2021mni}              & \cite{Ebert:1997ev,Ebert:2006nz,Faustov:2014zva,Faustov:2019mqr} & \cite{Verma:2011yw} & \cite{Cheng:2003sm} & \cite{Ivanov:2015tru,Ivanov:2017hun,Ivanov:2019nqd} & \cite{Wu:2006rd}
\\
\hline
$ B \rightarrow D $     &  0.70   & 0.67  &     0.6969               & 0.63   & $0.67(1)$ & $0.67$ & 0.78   &   \\
$ B \rightarrow \pi $   &  0.26   & 0.25  &                          & 0.217  &  0.25     &  0.25  & 0.28    & $0.285^{+0.016}_{-0.015}$ \\
$ D \rightarrow K $     &  0.75   & 0.79  & $0.744^{-(22)}_{+(23)}$  & 0.716  &  0.79(1)  &  0.78  & $0.77\pm 0.11$ & $0.661^{+ 0.067}_{-0.066}$ \\
$ D \rightarrow \pi  $ &  0.64   & 0.66  & $0.613^{-(21)}_{+(22)}$   & 0.640  &  0.66(1)  &  0.67  & $0.63\pm 0.09 $ & $0.635^{+0.060}_{-0.057}$ \\
\hline
\hline
\end{tabular}
\label{comparisons}
\end{table}
\end{center}
\end{widetext}

\hfill\break

Since we are not aware of experimental data away from the zero-recoil point to which we could compare our predictions, we make a comparison with lattice data which are given in Ref.~\cite{Abada:2000ty} for different values of $q^2$. This is done in Tab.~\ref{latticetable}. In most cases our predictions are within the statistical errors given for the lattice data. Actually the uncertainties in lattice calculations seem to be a little bit larger, as can be seen from Fig.~7 of Ref.~\cite{Hashimoto:2004hn}, where different lattice calculations, including Ref.~\cite{Abada:2000ty}, are compared for the $B\rightarrow \pi$ transition form factors. For a comprehensive and actual discussion of the problems connected with the determination of weak $B$ and $D$ transition form factors on the lattice we refer to the FLAG Review 2019~\cite{FlavourLatticeAveragingGroup:2019iem}.

\begin{table}[h!]
\begin{tabular}{ccc}
\multicolumn{3}{c}{\textbf{$B\rightarrow\pi$}} \\
\hline
\hline
$q^2$& Lattice\cite{Abada:2000ty}& IMF (this work)\\
\hline
13.6&$F_1=0.70(9)^{+.10}_{-.03}$  & $F_1=0.71 $ \\
    &$F_0=0.46(7)^{+.05}_{-.08}$  & $F_0=0.42 $ \\
\hline
15.0&$F_1=0.79(10)^{+.10}_{-.04}$ & $F_1=0.82 $ \\
    &$F_0=0.49(7)^{+.06}_{-.08}$  & $F_0=0.44 $ \\
\hline
17.9&$F_1=1.05(11)^{+.10}_{-.06}$ & $F_1=1.15 $ \\
    &$F_0=0.59(6)^{+.04}_{-.10}$  & $F_0=0.51 $ \\
\hline
20.7&$F_1=1.53(17)^{+.08}_{-.11}$ & $F_1=1.75 $ \\
    &$F_0=0.71(6)^{+.03}_{-.10}$  & $F_0=0.59 $ \\

\hline
\end{tabular}
\quad
\begin{tabular}{ccc}
 && \\
\multicolumn{3}{c}{\textbf{$D\rightarrow \pi$}} \\
\hline
\hline
$q^2$& Lattice \cite{Abada:2000ty}& IMF (this work)\\
\hline
0.47&$F_1=0.67(6)^{+.01}_{-.00}$  & $F_1=0.74 $ \\
    &$F_0=0.62(6)^{+.02}_{-.00}$  & $F_0=0.67 $ \\
\hline
0.97&$F_1=0.81(7)^{+.02}_{-.00}$ & $F_1=0.88 $ \\
    &$F_0=0.70(6)^{+.01}_{-.00}$  & $F_0=0.71 $ \\
\hline
1.48&$F_1=1.03(9)^{+.01}_{-.00}$ & $F_1=1.07 $ \\
    &$F_0=0.80(6)^{+.01}_{-.00}$  & $F_0=0.75 $ \\

\hline

 &    &  \\
   &   &  \\
      &   &  \\
%   \hline
\end{tabular}
\quad
\begin{tabular}{ccc}
&& \\
\multicolumn{3}{c}{\textbf{$D\rightarrow K$}} \\
\hline
\hline
$q^2$& Lattice \cite{Abada:2000ty}& IMF (this work)\\
\hline
0.19&$F_1=0.70(5)(0)$  & $F_1=0.79 $ \\
    &$F_0=0.68(4)(0)$  & $F_0=0.76 $ \\
\hline
0.69&$F_1=0.84(5)(0)$  & $F_1=0.91 $ \\
    &$F_0=0.76(4)(0)$ & $F_0=0.79 $ \\
\hline
1.7&$F_1=1.29(7)(0)$   & $F_1=1.27 $ \\
   &$F_0=0.96(4)(0)$   & $F_0=0.84 $ \\

\hline
 &    &  \\
   &   &  \\
      &   &  \\
\end{tabular}
\caption{Numerical values for $F_1$ and $F_0$ obtained by analytic continuation of the IMF results in comparison with lattice data \cite{Abada:2000ty}.}
\label{latticetable}
\end{table}

The agreement with other model calculations and with lattice data gives us some confidence that analytic continuation of our space-like transition form factors, calculated in the IMF, is an appropriate procedure to end up with a physically sensible model for the transition form factors in the time-like domain. It still remains to be seen, whether the difference between the  analytically continued IMF result and the direct decay calculation can be mainly attributed to a missing non-valence $Z$-graph contribution in the direct decay calculation, as often asserted. This question will be dealt with in  the sequel.

\subsection{$Z$-graph and meson pole}\label{sec:Zgraph}

\begin{figure}[h!]
\includegraphics[scale=0.4]{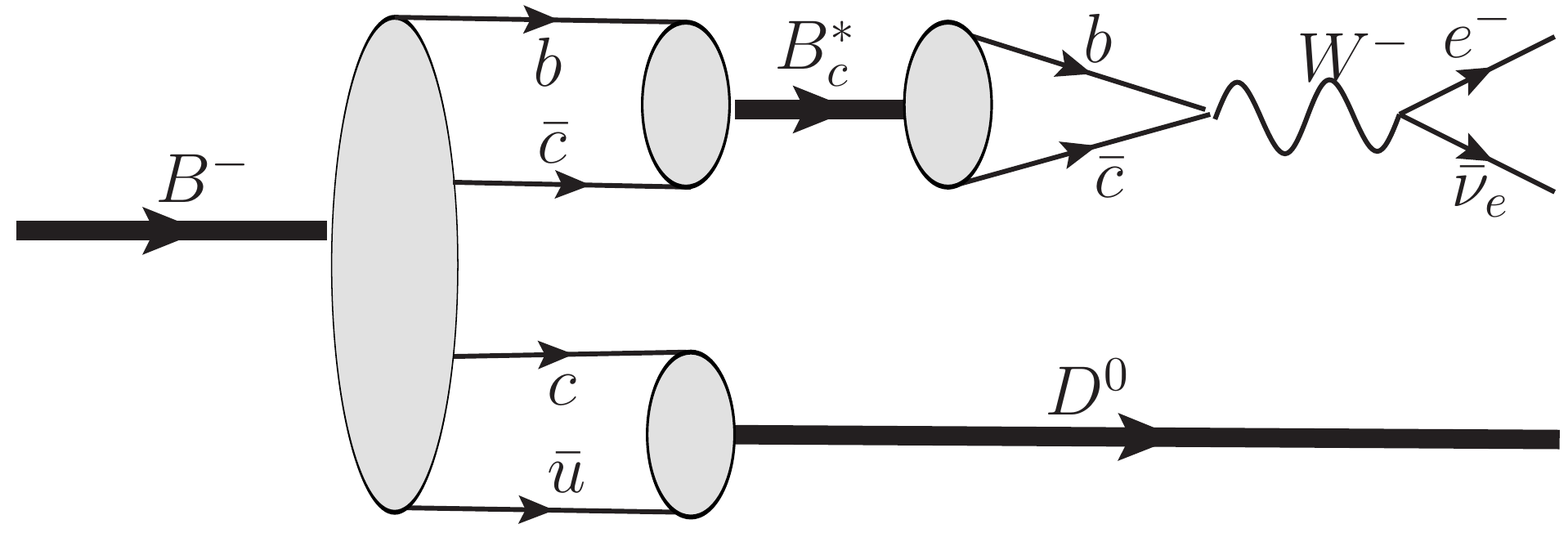}
\caption{Approximation of the non-valence $Z$-graph contribution to the $B^-\rightarrow D^0 e^- \bar{\nu}_e$ decay by means of a vector-meson-dominance-like mechanism.}
\label{fig:VMD}
\end{figure}

As already discussed in the introduction, the $Z$-graph contribution is well approximated by a vector-meson-dominance-like decay mechanism. As a consequence, the form factors exhibit a pole at an unphysical value of $q^2>0$  which should be observable  as an enhancement of the form factors near the zero-recoil point $q^2_{\mathrm{max}}$. Due to the confining forces, the non-valence degrees of freedom have to recombine with the valence $Q\bar q$ pair to color singlet hadrons. In the simplest case, they recombine to the outgoing meson $M^\prime$ and a vector-meson $M^\ast$ which fluctuates into a $W$ that decays in the sequel into $e\, \bar{\nu}_e$. This is depicted in Fig.~\ref{fig:VMD} for the $B^-\rightarrow D^0 e^- \bar{\nu}_e$ decay.

In the literature, the meson-pole contribution is included in different ways. In Ref.~\cite{Isgur:1989qw} the valence contribution and the $\bar{B}^{0\ast}$ pole contribution to the $\bar{B}^0\rightarrow \pi^+ e^- \bar{\nu}_e$ decay are calculated within a non-relativistic constituent-quark model. Thereby the $\bar{B}^0 \bar{B}^{0\ast} \pi^+$ transition is modeled by means of a quark-pair-creation mechanism~\cite{LeYaouanc:1972vsx} leading to a soft form factor at the $\bar{B}^{0\ast}\bar{B}^0 \pi$ vertex which suppresses the $\bar{B}^{0\ast}$ pole contribution away from the zero-recoil point. Reference~\cite{Cheung:1996qt}  uses relativistic front-form dynamics to analyze the $\bar{B}^0\rightarrow \pi^+ e^- \bar{\nu}_e$ decay. The valence contribution is described by means of a constituent-quark model, whereas the $B^\ast$-pole contribution is treated on hadron level with a phenomenological form factor at the $B^\ast B \pi$ vertex. In this paper also renormalization of the valence Fock state due to the presence of the $B^\ast D$ non-valence component is taken into account. The authors found that the valence contribution to the form factor $F_1$ in the low $q^2=0$ regime is well approximated by a function of the form:
\begin{eqnarray}
F_1^{\text{pole}}(q^2)
\es
{F_1(0) \over \left(1- {q^2\over m^2_{\text{pole}}} \right)^\alpha}\, ,
\label{pole}
\end{eqnarray}
with $\alpha=1.6$ amd $m_{\text{pole}}=5.32$ GeV. Towards the zero-recoil point the valence contribution to $F_1$ starts to decrease and deviates from this parameterization. Near zero recoil the non-valence $B^\ast$ pole contribution begins to dominate. The combined valence and $Z$-graphs results are also reasonably well approximated in the whole range $0\leq q^2 \leq q^2_{\mathrm{max }}$ by means of Eq.~(\ref{pole}) with parameters $\alpha=2.0$ and $m_{\text{pole}}=5.6-5.8$ GeV, depending on the strength of the $B^\ast B\pi$ coupling.

\begin{figure*}[t!]
\includegraphics[width=0.45\textwidth]{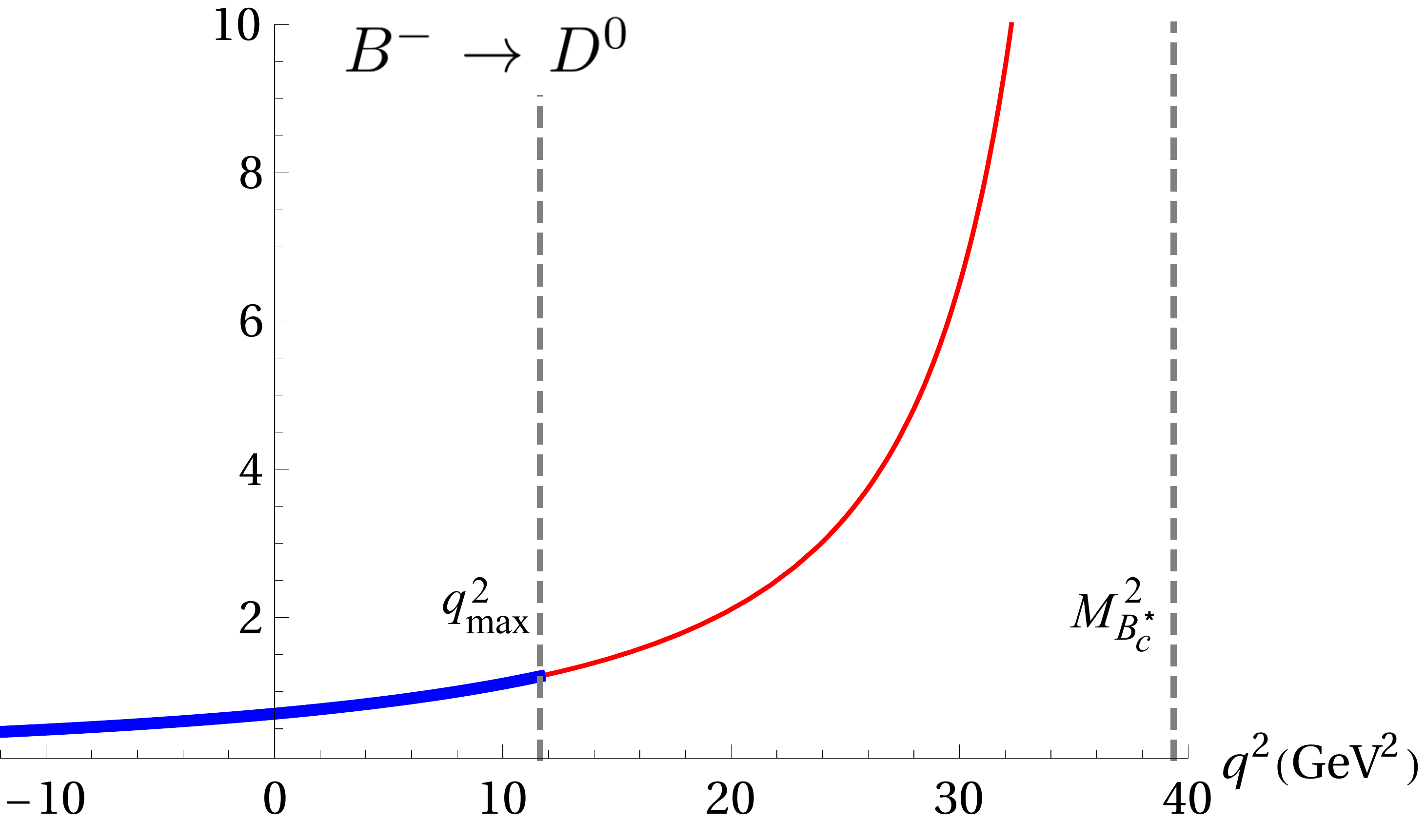}
\includegraphics[width=0.45\textwidth]{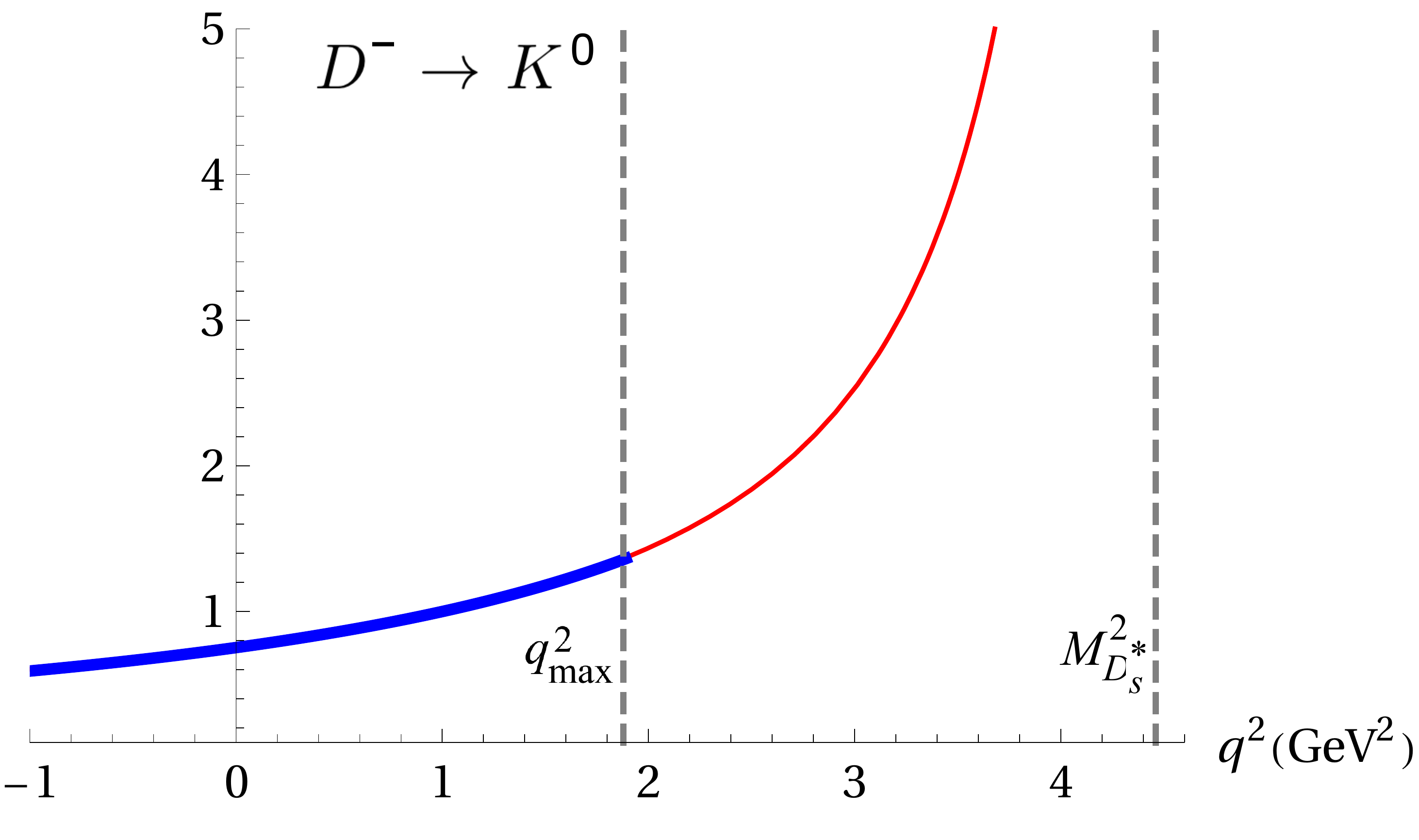}
\caption{Transition form factor $F_1$ for space- and timelike momentum transfers obtained by analytic continuation of the IMF result (thick, blue line) and pole fit (thin, red line) with $\alpha_{B\to D}=1.55$, $m_{\text{pole}}=m_{B_c^\ast}\approx m_{B_c}=6.274$ GeV (left) and $\alpha_{D\to K}=1.09$, $m_{D_s^{\ast}}=2.112$ GeV. The dashed, vertical lines indicate the position of $q^2_{\mathrm{max}}$ and $m^2_{\text{pole}}$.}
\label{ZBtoD-pole}
\end{figure*}

Following Ref.~\cite{Cheung:1996qt} we have also tried to parametrize our IMF results by means of Eq.~(\ref{pole}). But we have rather fixed the pole mass to the lightest vector-meson mass $M^\ast$ and left the power $\alpha$ as a free parameter. The outcome of our attempt is shown Figure~\ref{ZBtoD-pole} for the $B^-\rightarrow D^0$ and the $D^-\rightarrow K^0$ transitions.
With parameters $\alpha_{B\to D}=1.55$ and $\alpha_{D\to K}=1.09$, we find that our analytically continued IMF results follow Eq.~(\ref{pole}) surprisingly well in the plotted kinematic range. The closer the pole is to the zero-recoil point, the more the form factor $F_1$ tends to exhibit a monopole-like behavior as one would expect from  a meson-pole contribution. The masses of the incoming- and outgoing pseudoscalar mesons for various decays are given in Tab.~\ref{masstable} together with the corresponding intermediate-state vector meson masses and the distance of the pole from the zero-recoil point. Looking at this table it is obvious that the $B^-\rightarrow D^0$ decay form factor deviates more from a pure monopole behavior than the $D^-\rightarrow K^0$ decay form factor.   One must also keep in mind that we parameterize the whole form factor and not just the pole contribution and that the valence contribution gains importance with increasing vector-meson mass. It is, however, very remarkable that the analytic continuation of the IMF results provides an enhancement of the time-like form factors near zero recoil which resembles the $q^2$ behavior of a vector-meson-dominated decay mechanism, although we started with pure valence degrees-of-freedom and no additional dynamical input was explicitly introduced during the analytic continuation procedure. How this can happen and whether the analytically continued IMF result for the form factors can really be understood as the sum of a direct decay and a decay via an intermediate vector-meson (cf. Fig.~\ref{fig:VMD}), calculated in the rest frame of the decaying particle,  requires some additional modeling of the $M^\ast M M^\prime$ vertex and will be the subject of future investigations.

\begin{widetext}
\begin{center}
\begin{table}[h!]
\caption{Meson and pole masses}
\begin{tabular}{ccccc}
\hline
\hline
Transition&Initial Meson&Final Meson&Meson Pole&$m_{M^\ast}^2-q^2_{\mathrm{max}}$\\
\hline
 $B^-\rightarrow D^0$ & $m_{B^-}=5.2793$ GeV & $m_{D^0}=1.869$ GeV & $m_{B_c^{\ast}}\approx6.274$ GeV&27.73 Gev$^2$ \\
 $\bar{B}^0\rightarrow \pi^+$ & $m_{\bar{B}^0}=5.2797$ GeV & $m_{\pi^+}=0.1396$ GeV & $m_{B^{\ast}}=5.325$ GeV&1.935 GeV$^2$ \\
 $\bar{B}_S^0\rightarrow K^+$ & $m_{\bar{B}_S^0}=5.367$ GeV & $m_{K^+}=0.4937$ GeV & $m_{B^{\ast}}=5.325$ GeV&4.607 Gev$^2$ \\
 $\bar{D}^0\rightarrow K^+$ & $m_{\bar{D}^0}=1.864$ GeV & $m_{K^+}=0.4937$ GeV & $m_{D_s^{\ast}}=2.112$ GeV& 2.583 GeV$^2$ \\
 $D^-\rightarrow \pi^0$ & $m_{D^-}=1.869$ GeV & $m_{\pi^0}=0.135$ GeV & $m_{D^{\ast}}=2.010$ GeV&  1.033 GeV$^2$\\
\hline
\hline
\end{tabular}
\label{masstable}
\end{table}
\end{center}
\end{widetext}

\section{The heavy-quark limit}\label{sec:HQL}
In this section we want to verify, that the analytic continuation of the IMF form factors to $q^2\geq 0$  obeys the heavy-quark-symmetry requirements in the heavy-quark limit. Heavy-quark symmetry implies that the dynamics of the heavy-light meson becomes independent of the flavor and the spin of the heavy quarks as the heavy-quark masses go to infinity~\cite{Neubert:1993mb}. As one consequence,  electromagnetic and weak hadron form factors (multiplied with appropriate kinematical factors) approach a universal function, the Isgur-Wise function~\cite{Isgur:1990yhj}. In Refs.~\cite{GomezRocha:2012zd,Gomez-Rocha:2013bga} it has already been demonstrated that the electromagnetic and weak-decay form factors of heavy-light pseudoscalar and vector mesons, calculated within our (pure valence-quark) approach, satisfy the heavy-quark-symmetry constraints. Interestingly, the heavy-quark limit removes the frame-dependence of the form factors for space-like momentum transfers, indicating that $Z$-graph contributions are suppressed in the heavy-quark limit. This is what one would expect on physical grounds, since the probability for a higher Fock state containing an infinitely heavy quark-antiquark pair is strongly suppressed as compared to the valence Fock state. If this is the case, our analytically continued IF result and the result from the direct decay calculation should also agree in the heavy-quark limit and coincide with the Isgur-Wise function. This is what we are going to show now.

\begin{figure}[t!]
\includegraphics[width=0.47\textwidth]{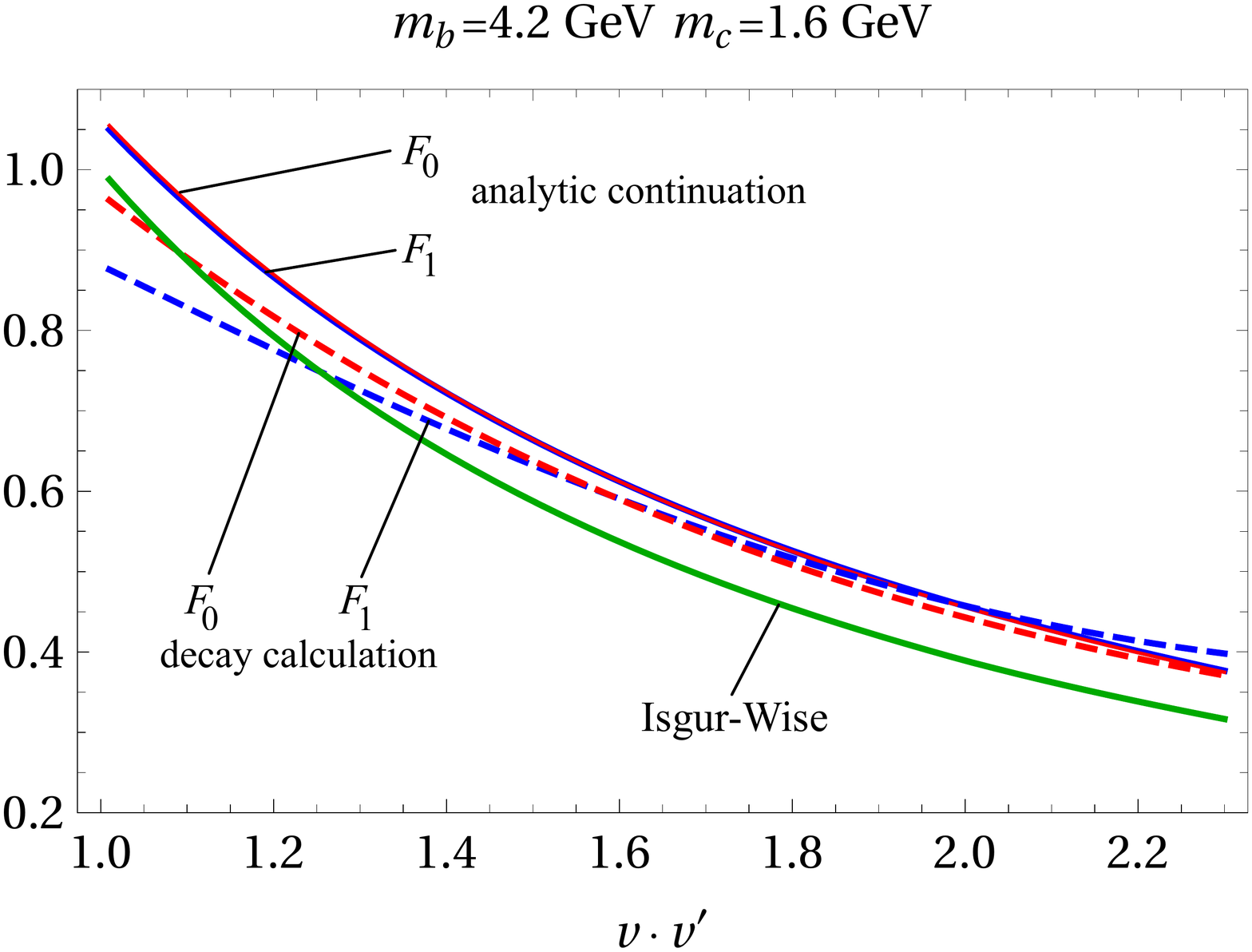}
\includegraphics[width=0.44\textwidth]{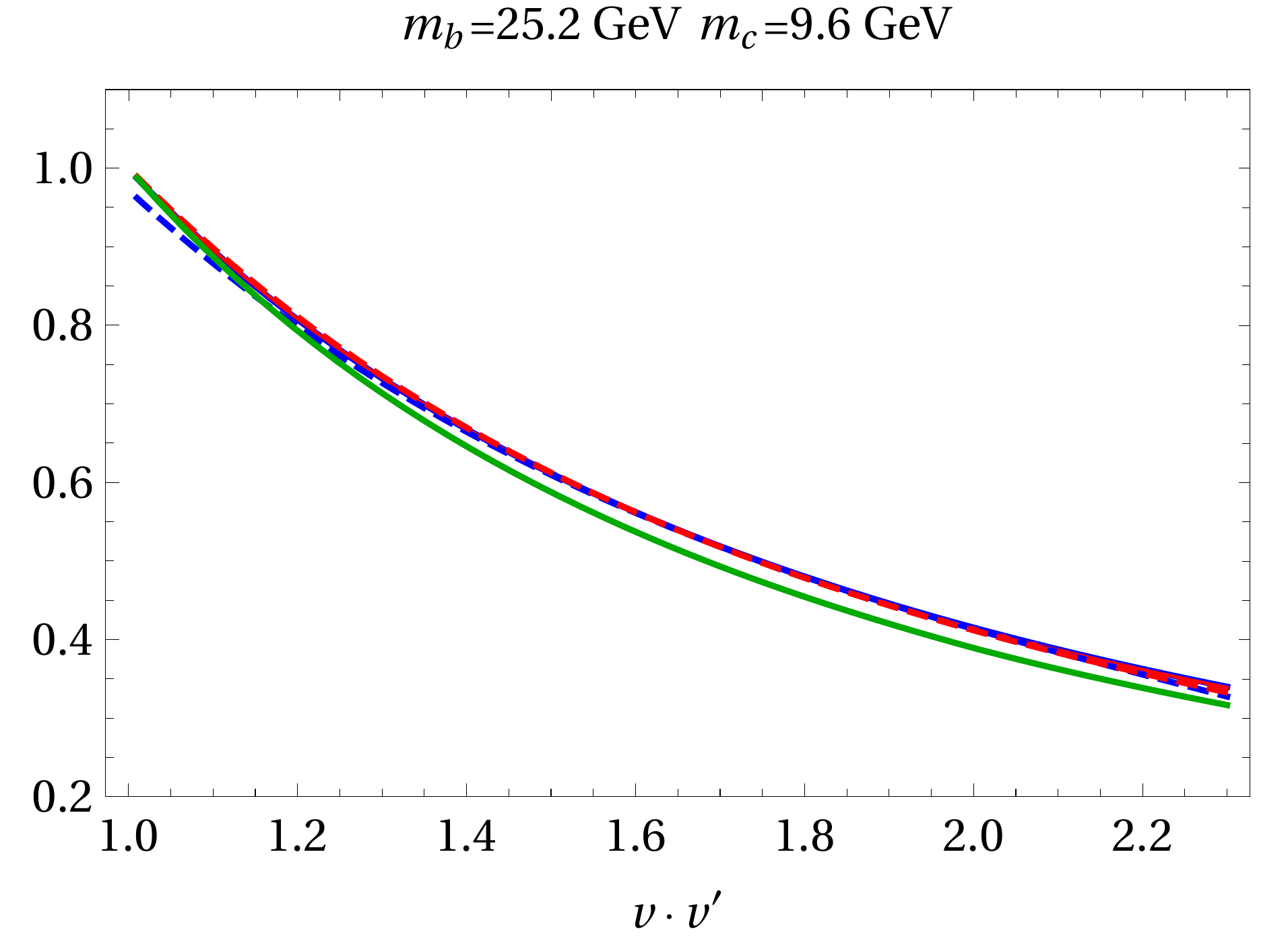}
\caption{ (Color online) $B^-\to D^0$ transition form factors as a function of $v\cdot v'$, multiplied by appropriate kinematical factors (see Eq.~(\ref{eq:FFIW}), in comparison with the Isgur-Wise function (green solid line). Upper panel: physical quark and meson masses. Lower panel: heavy quark masses are multiplied by a factor 6 and the meson masses are taken to be equal to the heavy-quark masses.}
\label{fig:IW}
\end{figure}

In order to perform the heavy-quark limit, the momentum transfer squared, $q^2=(p_M-p_{M^\prime})^2$, is first expressed in terms of
\begin{equation}
v_M\cdot v_{M^\prime}=\frac{p_M \cdot p_{M^\prime}}{m_M m_{M^\prime}}=\frac{m_M^2+m_{M^\prime}^2}{2 m_M m_{M^\prime}}-\frac{q^2}{2 m_M m_{M^\prime}}\, ,
\end{equation}
the product of the meson four-velocities. The heavy-quark limit $m_Q, m_{Q^\prime}\rightarrow\infty$ is then taken in such a way that $v_M\cdot v_{M^\prime}$ stays constant and both, the binding energy and the light-quark mass $m_q$, are neglected, i.e.
\begin{eqnarray}
m_{Q^{(\prime)}} \es m_{M^{(\prime)}} \ , \ \ {m_q \over m_{Q^{(\prime)}}} \ = \ 0 \ .
\end{eqnarray}
For a general discussion of the heavy-quark limit we refer to the review article~\cite{Neubert:1993mb}. Technical details on how the heavy-quark limit is is performed within our approach can be found in Refs.~\cite{GomezRocha:2012zd,Gomez-Rocha:2013bga}.
In the following we will focus on the $B^-\to D^0$ transition. As a consequence of heavy-quark symmetry both transition form factors, multiplied with appropriate kinematical factors, tend to one universal function, the Isgur-Wise function:
\begin{widetext}
\begin{eqnarray}\label{eq:FFIW}
R \left(1-\frac{q^2(v_B\cdot v_D)}{(m_b+m_D)^2}\right)^{-1} F_0\left(q^2(v_B\cdot v_D)\right) &\stackrel{m_b,m_c\rightarrow\infty}{\longrightarrow}& \xi(v_B\cdot v_D)\, ,\\
F_D(v_B\cdot v_D):=R\,F_1\left(q^2(v_B\cdot v_D)\right) \stackrel{m_b,m_c\rightarrow\infty}{\longrightarrow} \xi(v_B\cdot v_D)&&\hspace{-1.2cm}\hbox{with}\quad R=\frac{2 \sqrt{m_B m_D}}{m_B+m_D}\, .\nonumber
\end{eqnarray}
\end{widetext}
For physical quark and meson masses, deviations of these (rescaled) form factors from the Isgur-Wise function indicate the amount of heavy-quark-symmetry breaking. Figure~\ref{fig:IW} shows the Isgur-Wise function as resulting in the heavy-quark limit from the $B^-\rightarrow D^0$ decay form factors according to Eq.~(\ref{eq:FFIW}). As it turns out, it does not matter, whether one takes the form factors obtained from the direct decay calculation, or the analytically continued IMF form factors. This implies that the Isgur-Wise function is determined by the valence contribution and that cluster-separability violating effects vanish in the heavy-quark limit.  How the rescaled form factors tend to the Isgur-Wise function in the heavy-quark limit is also shown in Fig.~\ref{fig:IW} for both, the analytically continued IMF form factors and the form factors from the decay calculation. On the upper panel these form factors are plotted for physical quark and meson masses, on the lower panel these masses have been multiplied with a factor 6. For the upscaled masses one can already see that all form factors tend to the Isgur-Wise function, which means that heavy-quark symmetry is nearly restored. For physical masses of the heavy quarks the situation is somewhat different. The scaled form factors $F_0$ and $F_1$, as resulting from the analytic continuation, are nearly identical and coincide approximately with the Isgur-Wise function, if they are normalized to 1 at $v_B\cdot v_D=1$. This suggests that heavy-quark symmetry holds approximately already for physical quark masses and breaking effects just affect the normalization. On the other hand, for physical quark masses, the scaled form factors $F_0$ and $F_1$ from the direct decay calculation still differ from each other. Provided that the frame dependence of form factors can be mainly attributed to missing $Z$-graph contributions, this means that  their inclusion is also necessary to recognize approximate heavy-quark symmetry for physical quark masses.

In order to give a comparison with experimental data, we consider the slope of $F_1$ as a function for $v_M\cdot v_{M^\prime}$ at zero recoil. To be precise, the quantity we are interested in is defined as
\begin{eqnarray}
\rho^2_D &:=& - { F_1'(v_M\cdot v_{M^\prime}=1) \over F_1(v_M\cdot v_{M^\prime}=1) } \, ,
\end{eqnarray}
with $F^\prime_1$ meaning differentiation with respect to $v_M\cdot v_{M^\prime}$.
For the $B\to D$ decay  the experimental value provided by the heavy-flavor averaging group~\cite{HFLAV:2019otj} is
$\rho^2_D=1.131 \pm 0.033$. In~\cite{GomezRocha:2012zd,Gomez-Rocha:2013bga} the direct decay calculation involving only valence degrees of freedom gave a value of $\rho^2_D=0.55$ for physical quark masses and $\rho^2_D \stackrel{m_Q,m_{Q^\prime}\rightarrow\infty}{\longrightarrow} -\xi^\prime(1)=1.24$ in the heavy-quark limit. We confirm this finding. On the other hand, our analytically continued IMF form factor (for physical quark masses) has a value of $\rho^2_D=1.07$, much closer to the experiment than the direct decay calculation. This is a further support of the argument that the physically most complete description  of the $M\rightarrow M^\prime$ transition within a pure valence-quark picture is achieved in the IMF or, more generally, in a $q^+=0$ frame, if the calculation is done in front form. An equivalent description in other frames can most probably be achieved by adding the non-valence $Z$-graph contribution to the valence contribution.

\section{Conclusions and outlook}\label{sec:conclusions}

In this paper we have investigated weak $B\rightarrow D, \pi$ and $D\rightarrow K, \pi$ transition form factors for space- and time-like momentum transfers, as can be measured in neutrino scattering and semileptonic weak decays. We have used the point form of relativistic quantum mechanics in connection with the Bakamjian-Thomas construction to describe these systems in a relativistic invariant way by means of a constituent-quark model. The weak hadron transition current can then be extracted in a unique way from the invariant scattering or decay amplitude, which is calculated perturbatively in leading order of the weak coupling, starting from a multichannel mass operator. A covariant decomposition of this four-vector current yields the transition form factors.

For space-like momentum transfers $q^2<0$ these transition form factors were found to exhibit, in addition to the expected $q^2$ dependence, a dependence on the invariant mass squared $s$ of the scattering system. The same observation was already made in previous work on electromagnetic form factors~\cite{Biernat:2009my,GomezRocha:2012zd,Biernat:2014dea}, where it was suspected that the origin of this $s$-dependence are wrong cluster-separability properties inherent in the Bakamjian-Thomas construction. This $s$-dependence can be reinterpreted as a dependence on the frame in which the $WMM^\prime$ vertex is considered. Such a frame dependence is a common phenomenon and occurs, independent of the chosen form of relativistic dynamics, in any attempt to model a bound state current solely by means of valence Fock states using a one-body current. It is a common belief, which is supported by the analysis of the triangle diagram in a simple $\phi^3$ model, that this frame dependence can be cured by including a $Z$-graph contribution. Such a contribution requires, however, a non-valence component in the decaying meson. Fortunately, the non-valence component is suppressed in the infinite-momentum frame which corresponds in our approach to the limit $s\rightarrow \infty$. It is thus often asserted that the pure valence-quark picture provides already a physically satisfactory description of form factors, if these are calculated in the IMF. In any other frame the $Z$-graph can provide a non-negligible contribution to the form factors and should therefore be taken into account. If a missing $Z$-graph contribution is indeed responsible for the frame dependence of the form factors, its size in a particular frame will be approximately the difference between the form factors calculated in this frame and in the IMF.  Our numerical comparison of transition form factors $F_0$ and $F_1$ calculated in the Breit frame (i.e. $s$ minimal) with those calculated in the IMF (i.e. $s\rightarrow\infty$) showed that the differences are marginal for $B^-\rightarrow D^0, \pi^0$ transitions, whereas they can amount to about 10\% for ${D}^-\rightarrow K^0, \pi^0$ transitions.

The importance of the $Z$-graph contribution to the form factors is supposed to increase for time-like momentum transfers $q^2>0$. The annihilation process of the emitted quark-antiquark system into a $W$ is dominated by intermediate vector-meson states which allows to approximate the $Z$-graph contribution by a vector-meson-dominance-like decay mechanism. The closer the pole of the lightest vector meson is to the physical region, the more important becomes the $Z$-graph as compared to the pure valence contribution. But in contrast to scattering, the decay kinematics does not allow for a frame in which the $Z$-graph is suppressed. The invariant mass of the whole system is fixed to the rest mass of the decaying particle. In order to estimate the $Z$-graph contribution for time-like momentum transfers we have thus used the property that the form factors are meromorphic functions of $q^2$ which can be analytically continued from $q^2<0$ to $q^2>0$.  Under the assumption that the pure valence-quark picture provides already a physically satisfactory description of the form factors in the IMF, proper analytic continuation should also give a reasonably complete account of the form factors at time-like momentum transfers. If this is the case, the difference between the analytically continued IMF form factors and the form factors obtained from the direct decay calculation provides again an estimate for the size of the $Z$-graph contribution in the time-like domain. As suspected, these differences are already noticeable for the $B^-\rightarrow D^0$ decay and become really appreciable for the $B^-\rightarrow\pi^0$ and $D^-\rightarrow K^0,\pi^0$ decays.

For our numerical studies we have used a harmonic-oscillator form for the quark-antiquark meson wave functions with the oscillator parameters taken from Ref.~\cite{Cheng:1997}, where these were fitted to reproduce the known meson decay constants. With this parameterization of the meson wave functions, the analytic continuation of the IMF form factors gave quite reasonable results in the time-like region, which are in good agreement with existing lattice data~\cite{Hashimoto:2004hn}. Furthermore, we have also verified that the constraints put on the form factors by heavy-quark symmetry are satisfied within our approach. In the heavy-quark limit all form factors, written as functions of $v_M\cdot v_{M^\prime}$ and multiplied with appropriate kinematical factors, tend to one universal function, the Isgur-Wise function. The heavy-quark limit eliminates any frame dependence so that it even does not matter, whether the heavy-quark limit is applied to the analytically continued IMF form factors, or those from the direct decay calculation. Since the $Z$-graph vanishes in the heavy-quark  limit, this is a further indication that the difference between the IMF result and the direct decay calculation can be mainly ascribed to a missing $Z$-graph contribution in the decay calculation. We have also observed that approximate validity of heavy-quark symmetry holds already for physical masses of the heavy quarks in case of the IMF form factors. This allowed us to calculate $\xi^\prime(1)$, i.e. the slope of the Isgur-Wise function at zero recoil, for physical quark masses. For the $B\rightarrow D$ our prediction agrees very well with the experimentally determined value. For the form factors resulting from the direct decay calculation one has to go far beyond the physical quark masses in order to recover heavy quark symmetry. This means that the $Z$-graph could also be important to unveil heavy-quark-symmetry properties already for physical quark and meson masses.

\begin{figure*}[t!]
\includegraphics[width=0.8\textwidth]{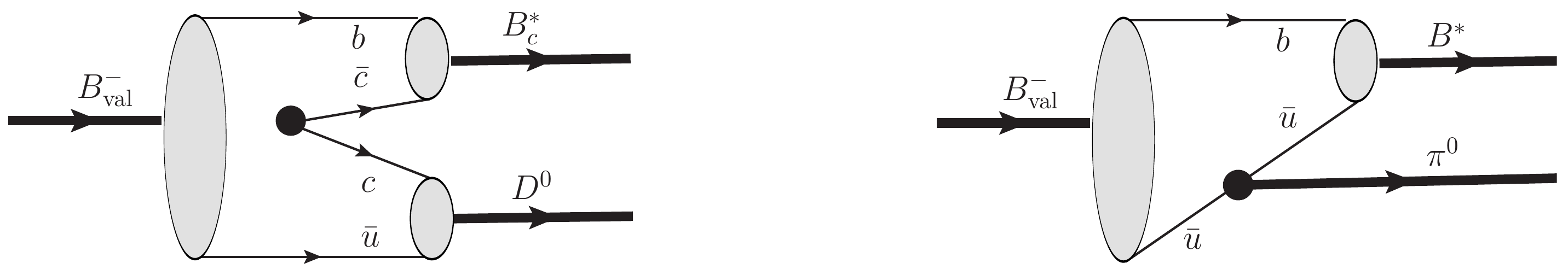}
\caption{Quark substructure of $M^\ast M M^\prime$ vertex in the $^3P_0$ model (left) and the chiral constituent-quark model (right). Blobs represent quark-antiquark wave functions for the respective mesons.}
\label{fig:3P0}
\end{figure*}

In this paper we have only tried to estimate the size of the $Z$-graph contribution indirectly. We have argued that the valence-quark picture gives a nearly complete description of the form factors in the IMF, since the $Z$-graph is suppressed in the IMF. If one goes to another frame, the observed deviation from the IMF result was then attributed to the missing $Z$-graph. This assumption was supported by the observation that $F_1(q^2)$ for the $B^-\rightarrow D^0$ and ${D}^-\rightarrow K^0$ decays is well approximated by a pole fit with the pole mass taken as the mass of the lightest intermediate vector meson and the power approaching a monopole behavior with decreasing vector-meson mass. Furthermore, the difference between the IMF result and the direct decay calculation vanishes in the heavy-quark limit, as one would expect from the $Z$-graph. But for a definite answer to the question, whether the IMF result provides indeed a complete physical picture and whether the frame dependence of the form factors, calculated within a pure valence-quark picture, can be cured by including a non-valence $Z$-graph contribution, the $Z$-graph has to be calculated explicitly. This would require some additional modeling, allowing for a non-valence component in the decaying meson.

The simplest way to model the dominant pole contribution, e.g. in the $B^-\rightarrow D^0$ decay, within a constituent-quark model is to assume that the physical $|B^-\rangle$ state is a superposition of the valence Fock state  $|B^-\rangle_{\mathrm{val}}$ and a non-valence $|D^0 B_c^\ast\rangle$ state, i.e. $|B^-\rangle=|B^-\rangle_{\mathrm{val}}+|D^0 B_c^\ast\rangle$. Such a model for the $B^-$ could, in principle, be accommodated within an extended constituent quark model which allows for the coupling of the valence $b\bar{u}$ channel to the non-valence $c\bar{u}b\bar{c}$ channel. For instantaneous confining forces between the quarks the corresponding mass eigenvalue problem can then be rewritten as a mass eigenvalue problem for purely hadronic degrees of freedom with the quark substructure just entering the strong interaction vertices~\cite{Kleinhappel:2012zj}. After an appropriate truncation of the sum over intermediate states one will end up with the simple picture of $B^-$ being the sum of a bare $B^-$ (i.e. the valence contribution) and a non-valence $D^0 B^\ast_c$ state. This way of including non-valence contributions has already been pursued for the (non-perturbative) calculation of hadron decay widths~\cite{Kleinhappel:2012zj,Schmidt:2017man} and pion-cloud effects in electromagnetic baryon form factors~\cite{Kupelwieser:2015ioa,Jung:2018lmk}. The new piece in our case would be a vertex that provides the transition from the $b\bar{u}$ channel to the $c\bar{u}b\bar{c}$ channel or, rephrased in terms of hadronic degrees of freedom, the $B_c^\ast B^- D^0$ vertex which has to be expressed in terms of quark degrees-of-freedom. Since a $c\bar{c}$ pair has to be created, a possible candidate for the vertex interaction would, e.g., be a $^3P_0$ model~\cite{Segovia:2012cd}, in which a quark-antiquark pair is created out of the vacuum. Decays with a $\pi$ or $K$ in the final state could be described even simpler within a chiral constituent-quark model~\cite{Glozman:1997ag} with the elementary degrees of freedom being constituent quarks and the octet of the lightest pseudoscalar mesons which couple directly to the constituent quarks. How the substructure of $M^\ast M M^\prime$ vertex would then look like in either the $^3P_0$ model, or the chiral constituent-quark model, is graphically represented in Fig.~\ref{fig:3P0} for the $B^\ast_c B^-_{\mathrm{val}}D^0$ and the $B^\ast B^-_{\mathrm{val}} \pi^0$ vertex, respectively. After having determined the $M^\ast M M^\prime$ vertex from its quark substructure one only has to solve a two-channel mass eigenvalue problem in which the  valence component $|M_{\mathrm{val}}\rangle$ is coupled to the non-valence component $|M^\prime M^\ast\rangle$. This gives the mass of the decaying meson $|M\rangle$, the probability of the valence component and the (two-particle) wave function of the non-valence component. With these ingredients, the $Z$-graph contribution to the weak transition form factors, in particular its relative importance as compared to the pure valence contribution, is determined uniquely from an underlying (extended) constituent-quark model. This kind of strategy has already been pursued within a front form approach in Ref.~\cite{Cheung:1996qt}, but there the $M^\ast M M^\prime$ vertex has not been determined from an underlying quark model, but rather parameterized in a purely phenomenological way. The explicit calculation of the $Z$-graph contribution to meson transition form factors in a way consistent with the underlying constituent-quark model, as we have just sketched, will be the subject of future work.

At the very end of the paper we still want to comment on the violation of cluster-separability within our approach. Mass operators with good cluster properties could, in principle, be constructed from our  mass operators (\ref{eq:mscatt}) and (\ref{eq:massopdecay}) by applying a series of unitary transformations~\cite{Sokolov:1977ym}. Formally it is known how to construct these unitary transformations~\cite{Coester:1982vt,Mutze:1984vw}, but the technical difficulties become tremendous when going beyond three-body systems and, in particular, for coupled-channel problems~\cite{Keister:1991sb}. To our knowledge, matrix elements of these unitary transformations have only been worked out explicitly for a three-body system consisting of distinguishable scalar particles~\cite{Kli:2018}, but we are not aware of any numerical studies which explicitly show how observables of a physical system change, when these unitary operators are applied~\cite{Keister:2011ie}. These unitary transformations introduce many-body forces and will add an effective many-body current to the one-body meson transition current we have used up till now. We have thus two sources for deviations of our model form factors from the frame independent physical form factors one wants to end up with: wrong cluster properties of the mass operator and a missing $Z$-graph contribution. The latter vanishes in the IMF and investigations on electromagnetic form factors have shown that the effect of wrong cluster properties is also minimized in the IF~\cite{Biernat:2010tp,Biernat:2014dea}. In this sense, form factors calculated in the IMF should be closest to the physical form factors we want to understand. Differences between form factors calculated in the IMF and any other frame receive, in principle, contributions from the $Z$-graph and from cluster-separability violating effects. In the course of this paper we have assumed that these differences are dominated by the $Z$-graph contribution, in order to estimate its size. This assumption is supported by the pole-like form factor behavior in the time-like region. But actually one should disentangle both contributions. To this aim it is at least necessary to determine the $Z$-graph contribution explicitly. If the frame dependence of the form factors, consisting of valence and $Z$-graph contribution, should then be negligible, it is most likely that cluster-separability-violating effects play a minor role like, e.g., asserted for nuclear systems~\cite{Keister:2011ie}. If this is not the case, it would be unavoidable to eliminate cluster-separability-violating effects by starting from mass operators with correct cluster properties.

\acknowledgements{M.G.R has been supported by the Spanish MINECO's Juan de la Cierva-Incorporaci\'on programme, Grant Agreement No. IJCI-2017-31531, Junta de Andaluc\'ia FQM-225, and project PID2020-114767GB-I00 funded by MCIN/AEI/10.13039/501100011033.}

\appendix

\section{Bound-state wave function and model parameters}
\label{app:wf}

To calculate the bound-state currents, cf.~Eq.(\ref{eq:Jwkpsps}) and~Eq.(\ref{eq:Jspsps}),  we use a simple harmonic-oscillator meson wave function:
\begin{eqnarray}
\psi_\alpha (\kappa) \es {2 \over \pi^{1\over 4} a_\alpha^{3\over 2}} e^{- {\kappa^2 \over 2 a_\alpha^2}} \ ,
\label{eq:wf}
\end{eqnarray}
where the subscript $\alpha$ distinguishes the different mesons.
We adopt the harmonic-oscillator parameters $a_\alpha$ and the quark constituent masses from Ref.~\cite{Cheng:1997}, where these parameters were fitted to reproduced meson decay constants. They are collected in Tab.~\ref{modelparameters}.

\begin{table}[h]
\centering
\caption{Model parameters.}
\label{modelparameters}
\begin{tabular}{lcr}
\hline
\hline
  $m_B=5.2793$ GeV & $m_b=4.8$ GeV  & $a_B=0.55$ GeV  \\
  $m_D=1.869$ GeV & $m_c=1.6$ GeV &$a_D=0.46$ GeV \\
  $m_{\pi}=0.1396$ GeV & $m_{u,d}=0.25$ GeV & $a_{\pi}=0.33$ GeV \\
  $m_K=0.4937$ GeV & $m_s=0.4$ GeV & $a_K=0.38$ GeV \\
\hline
\hline
\end{tabular}
\end{table}

\bibliography{PF}{}

\bibliographystyle{apsrev4-1}

\end{document}